\documentclass[prb,twocolumn,showpacs]{revtex4}
\usepackage{graphicx}
\begin{document}
\date{\today}
\title{BCS-BEC crossover at finite temperature in the 
broken-symmetry phase}
\author{P. Pieri, L. Pisani, and G.C. Strinati}
\affiliation{Dipartimento di Fisica, UdR INFM,
Universit\`{a} di Camerino, I-62032 Camerino, Italy}

\begin{abstract}
The BCS-BEC crossover is studied in a systematic way in the 
broken-symmetry phase between zero temperature and the critical temperature. 
This study bridges two regimes where quantum and thermal fluctuations are, 
respectively, important.
The theory is implemented on physical grounds, by adopting a fermionic 
self-energy in the broken-symmetry phase that represents fermions coupled to
superconducting fluctuations in weak coupling and to bosons described
by the Bogoliubov theory in strong coupling.
This extension of the theory beyond mean field proves 
important at finite temperature, to connect with the results in the
normal phase. The order parameter, the chemical 
potential, and the single-particle spectral function are calculated numerically
for a wide range of coupling and temperature. This enables us to assess the 
quantitative importance of superconducting fluctuations in the broken-symmetry
phase over the whole BCS-BEC  crossover. Our results are 
relevant to the possible realizations of this crossover with high-temperature 
cuprate superconductors and with ultracold fermionic atoms in a trap. 
\end{abstract}


\maketitle

\section{Introduction}
In the BCS to Bose-Einstein condensation (BEC) crossover
\cite{Eagles-69,Leggett-80,NSR-85,Randeria-90,Haussmann-93,PS-94,PS-96,Levin-97,Zwerger-97,Pi-S-98}, 
largely overlapping Cooper pairs
smoothly evolve into non-overlapping composite bosons as the fermionic 
attraction is progressively increased. These two physical situations 
(Cooper pairs vs composite bosons) correspond to the weak- and strong-coupling
limits of
the theory, while in the interesting intermediate-coupling regime neither
the fermionic nor the bosonic properties are fully realized. 
Under these circumstances,
the theory is fully controlled on the weak- and strong-coupling sides,
while at intermediate coupling an interpolation scheme results (as for all
crossover approaches).
These physical ideas are implemented, in practice, by allowing for a strong 
decrease of the chemical potential {\em at a given temperature} when passing 
from the weak- to the strong-coupling limit. 

The BCS-BEC crossover can be 
considered both below (broken-symmetry phase) and above (normal phase) the
superconducting critical temperature. In particular, in the normal phase 
preformed pairs exist in the strong-coupling limit up to a temperature 
$T^*$ corresponding to the breaking of the pairs, while coherence among the 
pairs is established when the temperature is lowered below the superconducting 
critical temperature $T_c$. This framework could be relevant 
to the evolution of the properties of high-temperature cuprate superconductors
from the overdoped (weak-coupling) to the underdoped (strong-coupling) regions
of their phase diagram~\cite{Damascelli}. The BCS-BEC crossover can be also 
explicitly 
realized with ultracold fermionic atoms in a trap, by varying their mutual 
effective attractive interaction via a Fano-Feshbach resonance~\cite{expcross}.

The BCS-BEC crossover has been studied extensively in the past, either at 
$T=0$ or for $T\ge T_c$.
At $T=0$, the solution of the two coupled BCS (mean-field) equations for the 
order parameter $\Delta$ and the chemical potential $\mu$ has been 
shown to cross over smoothly from a BCS weak-coupling 
superconductor with largely overlapping Cooper pairs to a strong-coupling 
superconductor where tightly-bound pairs are condensed in a Bose-Einstein
(coherent) ground state~\cite{Eagles-69,Leggett-80,MPS}. 
For this reason, the BCS mean field has often been considered
to be a reliable approximation for studying the whole BCS-BEC crossover at 
$T=0$.
At finite temperature, the increasing importance in 
strong coupling of the thermal excitation of collective modes 
(corresponding to noncondensed bosons) was first pointed out by 
Nozi\`eres and Schmitt-Rink~\cite{NSR-85}.
By their approach, the expected result that the superconducting critical 
temperature should approach the Bose-Einstein temperature $T_{{\rm BE}}$ in 
strong coupling was obtained (coming from {\em above $T_c$}) via a 
(first-order) inclusion of the $t$-matrix self-energy in the fermionic 
single-particle Green's function.
The same type of $t$-matrix approximation (also with the inclusion, by some 
authors, of self-consistency) has then been widely adopted to study  
the BCS-BEC crossover above $T_c$, both for continuum~\cite{Haussmann-93} 
and lattice models~\cite{Fresard,Micnas,Randeria-97-1,KKK}.

Despite its conceptual importance, a systematic study of the BCS-BEC crossover 
in the temperature range $0<T<T_c$ is still lacking.
A diagrammatic theory for the BCS-BEC crossover that extends 
below $T_c$ the self-consistent $t$-matrix approximation was proposed some 
time ago by Haussmann~\cite{Haussmann-93}. The ensuing coupled equations for
the order parameter and chemical potential were,
however, solved explicitly only at $T_c$,~\cite{Haussmann-2} leaving therefore 
unsolved the problem of the study of the whole temperature 
region below $T_c$. The work by Levin and coworkers~\cite{levin98}, on the 
other hand, even though based on a ``preformed-pair scenario'', has focused
mainly on the weak-to-intermediate coupling region, where the 
fermionic chemical potential remains inside the single-particle band.
An extension of the self-consistent 
$t$-matrix approximation to the superconducting phase for a two-dimensional 
lattice model was considered in Ref.~\onlinecite{Yanase-00}.
In that paper, however, the shift of the chemical potential associated 
with the 
increasing coupling strength was ignored, by keeping it fixed at the
noninteracting value.\cite{footnote-1}
The results of Ref.~\onlinecite{Yanase-00}  
are thus not appropriate to address the BCS-BEC crossover, for which the 
renormalization of the chemical potential (that evolves from the 
Fermi energy in weak coupling to half the binding energy of a pair in 
strong coupling) plays a crucial 
role~\cite{Eagles-69,Leggett-80,NSR-85}. Additional studies have made use of
a fermion-boson model~\cite{ranninger}, especially in the context of trapped
Fermi gases~\cite{ohashi}.

Purpose of the present paper is to study the BCS-BEC crossover in the 
superconducting phase over the whole temperature range from $T=0$ to $T=T_c$, 
thus filling a noticeable gap in the literature. 
We will consider a three-dimensional continuum model, for which the fermionic 
attraction can be modeled by a point-contact interaction.
As noted in Refs.~\onlinecite{Haussmann-93} and~\onlinecite{Pi-S-98}, with 
this model
the structure of the diagrammatic theory for the single-particle fermionic 
self-energy simplifies considerably, since only limited 
sets of diagrammatic structures survive the regularization of the contact 
potential in terms of the fermionic two-body scattering length 
$a_F$.~\cite{Randeria-93,Pi-S-98}
The dimensionless interaction parameter $(k_{F} a_{F})^{-1}$ (where the
Fermi wave vector $k_{F}$ is related to the density 
via $n=k_{F}^{3}/(3\pi^{2})$) then ranges from $-\infty$ in weak coupling
to $+\infty$ in strong coupling. The crossover region of interest is, however,
restricted in practice by $(k_{F} |a_{F}|)^{-1} \lesssim 1$.

For this model, a systematic theoretical study of the evolution of the 
single-particle spectral function in the normal phase from the BCS to BEC 
limits has been presented recently~\cite{PPSC-02}.
Like in Ref.~\onlinecite{NSR-85}, also in Ref.~\onlinecite{PPSC-02} the 
coupling of a fermionic single-particle 
excitation to a (bosonic) superconducting fluctuation mode was taken into 
account by the $t$-matrix self-energy. This approximation embodies the 
physics of a dilute  Fermi gas in the weak-coupling limit 
and reduces to a description of independent composite bosons in the 
strong-coupling limit.
In this way, single-particle spectra were obtained in 
Ref.~\onlinecite{PPSC-02} as functions of coupling strength and temperature.

In the present paper, the $t$-matrix approximation for the self-energy is 
suitably extended below $T_c$. 
In particular, the \emph{same\/} superconducting fluctuations, that in 
Refs.~\onlinecite{NSR-85} and \onlinecite{PPSC-02} were coupled to  
fermionic independent-particle excitations above $T_{c}$, are now coupled to 
fermionic BCS-like single-particle excitations below $T_{c}$.
In the strong-coupling limit, it turns out that these 
superconducting fluctuations merge in a nontrivial way\cite{APS-02} into a 
state of condensed composite bosons described by the
Bogoliubov theory, and evolve consistently into a state of 
independent composite bosons above $T_{c}$ (as the Bogoliubov theory for
point-like bosons does \cite{Bassani-GCS-01}).
In this way, a direct connection is established between the structures of
the single-particle fermionic self-energy above \emph{and\/} below $T_{c}$, 
as they embody the same kind of bosonic mode which itself
evolves with temperature. 

A comment on the validity of the Bogoliubov theory at finite temperature (and,
in particular, close to the Bose-Einstein transition temperature $T_{BE}$) 
might be relevant at this point.
A consistent theory for a \emph{dilute\/} condensed Bose gas was  
developed long ago in terms of a (small)
gas parameter \cite{Beliaev-58,Popov-87}, of which the Bogoliubov 
theory \cite{FW} is only an approximate form valid at
low enough temperatures (compared with $T_{BE}$).
That theory correctly describes also the dilute Bose gas in the normal phase
\cite{Popov-87}, whereas the Bogoliubov
theory (when extrapolated above the  critical temperature) recovers the
independent-boson form (albeit
in a non-monotonic way, with a discontinuous jump affecting the bosonic
condensate \cite{Bassani-GCS-01}).
It would therefore be desirable to
identify (at least in principle) a fermionic theory that, in
the strong-coupling limit of the fermionic attraction, maps onto a
more sophisticated bosonic theory, overcoming the apparent limitations of the 
Bogoliubov theory. In practice, however,  it should be considered already a
nontrivial achievement of the present approach
the fact that the bosonic Bogoliubov approximation can be reproduced 
from an originally fermionic theory.
For these reasons, and also because it is actually the intermediate-coupling
(crossover) region to be of most physical interest, in the following we shall 
consider the Bogoliubov approximation as a reasonable 
limiting form of our fermionic theory.

As it is always the case for the BCS-BEC crossover approach, implementation
of the theory developed in
the present paper rests on solving two coupled equations for the
order parameter $\Delta$ and the chemical potential $\mu$.
The equations here considered for $\Delta$ and $\mu$ generalize
the usual equations already considered at the mean-field 
level~\cite{Eagles-69,Leggett-80,NSR-85}, by including fluctuation 
corrections.
Our equations reproduce the expected physics in the strong-coupling limit,
at least at the level of approximation here considered. Their solution 
provides us with the values of $\Delta$ and $\mu$ as functions of coupling 
strength $(k_{F} a_{F})^{-1}$ and temperature $T$, thus extending 
results obtained previously at the mean-field level.
In particular, the order parameter is now found to vanish at a temperature
(close to) $T_{c}$ even in the strong-coupling limit, while it would had 
vanished close to $T^{*}$ at the mean-field level \cite{PE-00}.

The analytic continuation of the fermionic self-energy to the real
frequency axis is further performed
to obtain the single-particle spectral function $A({\mathbf k},\omega)$,
that we study in a systematic way
as a function of wave vector ${\mathbf k}$, frequency $\omega$, coupling
strength $(k_{F} a_{F})^{-1}$, and temperature $T$.
In this context, two novel sum rules (specific to the broken-symmetry phase)
are obtained, which provide compelling checks on the numerical 
calculations.
In addition, the numerical calculations are tested against analytic (or
semi-analytic) approximations obtained in the strong-coupling 
limit. The study of a dynamical quantity like $A({\bf k},\omega)$ enables
us to attempt a comparison with the experimental ARPES and tunneling spectra 
for 
cuprate superconductors below $T_c$, for which a large amount of data exists
showing peculiar features for different doping levels and temperatures. As in 
Ref.~\onlinecite{PPSC-02} above $T_c$, this comparison concerns especially
the experimental data about the M points in the Brillouin zone of cuprates,
where pairing effects are supposed to be stronger than along the nodal lines.

Our main results are the following. About thermodynamic quantities, we 
will show that fluctuation corrections over and above 
mean field are especially important at finite temperature 
$T\lesssim T_c$ when approaching the strong-coupling limit. At zero 
temperature, fluctuation corrections to thermodynamic 
quantities turn out to be of some relevance only in the intermediate-coupling 
region. This supports the expectation~\cite{Leggett-80} that the BCS mean 
field 
at zero temperature should describe rather well the BCS-BEC crossover
essentially for all couplings. Regarding instead dynamical quantities like 
$A({\bf k},\omega)$, our calculation based on a 
``preformed-pair scenario'' reveals two distinct spectral features for 
$\omega < 0$. These features, which have different temperature and doping 
dependences, together give rise to a peak-dip-hump structure which is 
actively debated 
for the ARPES spectra of cuprate superconductors. Our results differ 
from those previously obtained by other calculations~\cite{levin98} also based
 on a ``preformed-pair scenario'', where a single feature was instead
obtained in the spectral function for $\omega < 0$. An explanation of this
discrepancy between the two calculations will be provided.
 It will 
also turn out from our calculation that the coherent part of  
$A({\bf k},\omega)$ for $\omega < 0$ follows essentially a BCS-like behavior as
far as its wave-vector dependence is concerned, albeit with a gap value which
contains an important contribution from fluctuations at finite temperature.
The same BCS-like behavior is not found, however, by our calculation  for the 
dependence of the spectral weight of the coherent peak on temperature and 
coupling.
This evidences a dichotomy in the behavior of $A({\bf k},\omega)$, 
according to which of its dependences one is after. 
Such a dichotomy is clearly observed in experiments on cuprate 
superconductors, in good qualitative agreement with the results obtained by 
our calculations.~\cite{pps04}.  
A detailed quantitative comparison of our results with the experimental data 
on cuprates would, however, require a more refined theoretical model, as to 
include the quasi-two-dimensional lattice structure, the $d$-wave character 
of the superconducting gap, and also a fermionic attraction that depends 
effectively on doping (and possibly on temperature).
Future work on this subject should address these additional issues.

The present theory could be improved in several ways. In the present approach, 
the effective interaction between the composite bosons is treated within the 
Born approximation. For a dilute system of composite bosons one knows how
to improve on this result, as shown in Ref.~\onlinecite{Pi-S-98} (see also 
Ref.~\onlinecite{petrov}).
In addition, the Bogoliubov description for the composite bosons could be also
improved, for instance, by extending to the composite bosons the Popov 
treatment for point-like bosons~\cite{Popov-87}. Finally, on the weak-coupling
side of the crossover the BCS theory could be modified by including the 
contributions shown by Gor'kov and Melik-Barkhudarov~\cite{gmb} to yield a 
finite 
renormalization of the critical temperature and of the gap function 
{\em even\/} in the extreme weak-coupling limit. Work along these lines is in 
progress.

The plan of the paper is as follows.
In Sec.~II we discuss our choice for the fermionic self-energy in the
superconducting phase, from which
the order parameter $\Delta$ and the chemical potential $\mu$ are obtained
as functions of temperature and coupling strength, and the spectral 
function $A({\mathbf k},\omega)$ also results.
Analytic results are presented in the strong-coupling limit, where the
order parameter is shown to be connected with the bosonic condensate density 
of the Bogoliubov theory.
In addition, the analytic continuation of our expressions for the
fermionic self-energy and spectral function is carried out in detail.
In Sec.~III we present our numerical calculations, 
and discuss the results for the single-particle spectral function in the
context of the available experimental data for high-temperature cuprate
superconductors.
Section IV gives our conclusions.
In Appendix A two sum rules are derived for the superconducting phase, which 
are used as checks of the numerical results.
\section{Diagrammatic theory for the BCS-BEC crossover in the 
superconducting phase}

In this section, we discuss the choice of the fermionic single-particle 
self-energy in the superconducting
phase for a (three-dimensional) continuum system of fermions mutually
interacting via an attractive point-contact
potential, with an $s$-wave order parameter.
We shall place special emphasis to the strong-coupling limit of 
the theory, where composite bosons forms as bound fermion pairs.
We extend in this way \emph{below\/} $T_{c}$ an analogous 
treatment for the self-energy, made previously in the normal 
phase to calculate the single-particle spectral function.\cite{PPSC-02} 

Knowledge of the detailed form of the attractive interaction is not
generally required when studying the BCS-BEC
crossover.
Accordingly, one may consider the simple form 
$v_{0} \delta ({\mathbf r})$ of a ``contact'' potential,
where $v_{0}$ is a negative constant.
This choice entails a suitable regularization in terms, e.g., of a cutoff
$k_{0}$ in wave-vector space.
In three dimensions, this is achieved via the scattering length $a_{F}$ of
the associated fermionic two-body
problem, by choosing $v_{0}$ as follows~\cite{Pi-S-98}:
\begin{equation}
v_{0} \, = \, - \, \frac{2 \pi^{2}}{m k_{0}} \, - \,
           \frac{\pi^{3}}{m a_{F} k_{0}^{2}}  \,\,\, 
\label{v0}
\end{equation}
$m$ being the fermion mass. With this choice, the classification of the 
(fermionic) many-body diagrams is
considerably simplified not only in the
normal phase \cite{Pi-S-98} but also in the broken-symmetry phase
\cite{APS-02}, since only specific diagrammatic
substructures survive when the limit $k_{0} \rightarrow \infty$ (and thus 
$v_{0} \rightarrow 0$) is eventually taken.

In particular, the \emph{particle-particle ladder\/} depicted in Fig.~1(a)
survives the regularization of the
potential.\cite{footnote-Nambu-arrows} It is obtained by the matrix inversion:
\begin{eqnarray}
\left( 
\begin{array}{cc}
\Gamma_{11}(q)&\Gamma_{12}(q)\\
\Gamma_{21}(q)&\Gamma_{22}(q)\end{array}\right)
&=& 
\left( 
\begin{array}{cc} 
\chi_{11}(-q)&\chi_{12}(q)\\ 
\chi_{12}(q)&\chi_{11}(q)\end{array}\right)\nonumber\\
&\times& [\chi_{11}(q)  \chi_{11}(-q) - \chi_{12}(q)^{2}]^{-1}
\label{Gamma-solution}
\end{eqnarray}
with the notation
\begin{eqnarray}
- \chi_{11}(q) &=& \frac{m}{4\pi a_F} +  \int \! \frac{d {\mathbf p}}{(2\pi)^{3}} \left[
\frac{1}{\beta} \sum_{n} 
{\mathcal G}_{11}(p+q) {\mathcal G}_{11}(-p) 
\right.\nonumber \\
& &\phantom{\frac{m}{4\pi a_F}}\phantom{\frac{m}{4\pi a_F}}
\phantom{\frac{m}{4\pi a_F}} - \left.\frac{m}{|{\bf p}|^2}\right]
\label{A-definition}\\
\chi_{12}(q) & = & \int \! \frac{d {\mathbf p}}{(2\pi)^{3}} \,
\frac{1}{\beta} \, \sum_{n} \,
{\mathcal G}_{12}(p+q) \,{\mathcal G}_{21}(-p)           \,\,\, .
\label{B-definition}
\end{eqnarray}
In these expressions, $q=({\mathbf q},\Omega_{\nu})$ and
$p=({\mathbf p},\omega_{n})$, where
${\mathbf q}$ and ${\mathbf p}$ are wave vectors, and
$\Omega_{\nu}=2\pi\nu/\beta$ ($\nu$ integer) and
$\omega_{n}=(2n+1)\pi/\beta$ ($n$ integer) are bosonic and fermionic
Matsubara frequencies, respectively
(with $\beta=(k_{B}T)^{-1}$,  $k_{B}$ being the Boltzmann's constant);
\begin{eqnarray}
{\mathcal G}_{1 1}({\mathbf p},\omega_n) \, & = & \, - \frac{\xi({\mathbf p})
+ i \omega_n}
{E({\mathbf p})^2 + \omega_n^2} \, = \, - \, {\mathcal G}_{2
2}({\mathbf -p},-\omega_n)  \nonumber  \\
{\mathcal G}_{2 1}({\mathbf p},\omega_n) \, & = & \, \frac{\Delta}
{E({\mathbf p})^2 + \omega_n^2} \, = \, {\mathcal G}_{1
2}({\mathbf p},\omega_n)  
\label{BCS-Green-function}
\end{eqnarray}
are the BCS single-particle Green's functions in Nambu notation, with
$\xi({\mathbf p})={\mathbf p}^{2}/(2m) - \mu$
and $E({\mathbf p})=\sqrt{\xi({\mathbf p})^{2}+\Delta^{2}}$ for an isotropic
($s$-wave) order parameter $\Delta$.
[Hereafter, we shall take the order parameter to be real with no loss of
generality.]

The expressions (\ref{A-definition}) and (\ref{B-definition}) for 
$\chi_{11}(q)$ and $\chi_{12}(q)$ considerably simplify 
\emph{in the strong-coupling limit\/}
(that is, when $\beta \mu \rightarrow - \infty$ and $\Delta \ll |\mu|$).
In this limit, one then obtains for the matrix elements
(\ref{Gamma-solution}) \cite{Haussmann-93,APS-02}:
\begin{equation}
\Gamma_{11}(q) \, = \, \Gamma_{22}(-q) \, \simeq \, \frac{8 \pi}{m^{2}
a_{F}} \,
\frac{\mu_{B} \, + \, i \Omega_{\nu} \, + \, {\mathbf q}^{2}/(4m)}
{E_{B}({\mathbf q})^{2} \, - \, (i \Omega_{\nu})^{2}}
\label{Gamma-11-approx}
\end{equation}
and
\begin{equation}
\Gamma_{12}(q) \, = \, \Gamma_{21}(q) \, \simeq \, \frac{8 \pi}{m^{2} a_{F}} \,
\frac{\mu_{B}}{E_{B}({\mathbf q})^{2} \, - \, (i \Omega_{\nu})^{2}}  \,\,\,
,         \label{Gamma-12-approx}
\end{equation}
where
\begin{equation}
E_{B}({\mathbf q}) \, = \, \sqrt{ \left( \frac{{\mathbf q}^{2}}{2 m_{B}} \, +
\, \mu_{B}\right)^{2}
                                     \, - \, \mu_{B}^{2}}
\label{Bogoliubov-disp}
\end{equation}
has the form of the Bogoliubov dispersion relation \cite{FW} 
($m_{B}=2m$ being the bosonic mass,
 $\mu_{B} = \Delta^{2}/(4 |\mu|) = 2 \mu + \epsilon_{0}$ the bosonic
chemical potential, and
$\epsilon_{0} = (m a_{F}^{2})^{-1}$ the bound-state energy of the
associated fermionic two-body
problem).
The above relation between the fermionic and bosonic chemical potentials
holds provided $\mu_{B} \ll \epsilon_{0}$
(cf. also Sec.~IID).
Note that $\mu_{B}$ can be cast in the Bogoliubov form
\begin{equation}
\mu_{B} \, = \, v_{2}(0) \,\, n_{0}(T) \label{pot-chim-Bog}
\end{equation}
where $v_{2}(0)=4 \pi a_{F}/m$ is the residual bosonic interaction
\cite{Haussmann-93,Pi-S-98} and
$n_{0}(T)=\Delta^2(T) m^2 a_F/(8\pi)$ is the {\em condensate density\/}.
The relation (\ref{pot-chim-Bog}) is formally obtained already at the
(BCS) mean-field level
\cite{APS-02}, albeit with an unspecified dependence of $n_{0}(T)$ on
temperature. Within our fluctuation
theory, the temperature dependence of $n_{0}(T)$ will coincide in strong 
coupling with the expression given by the Bogoliubov theory (see Sec.~IID).
In particular, at zero temperature and at the lowest order in the residual
bosonic interaction\cite{APS-02}, $n_{0}$
reduces to the bosonic density $n_{B}=n/2$ and $\mu_{B}$ is given by $2
k_{F}^{3} a_{F}/(3 \pi m)$.

\begin{figure}
\includegraphics[scale=0.35]{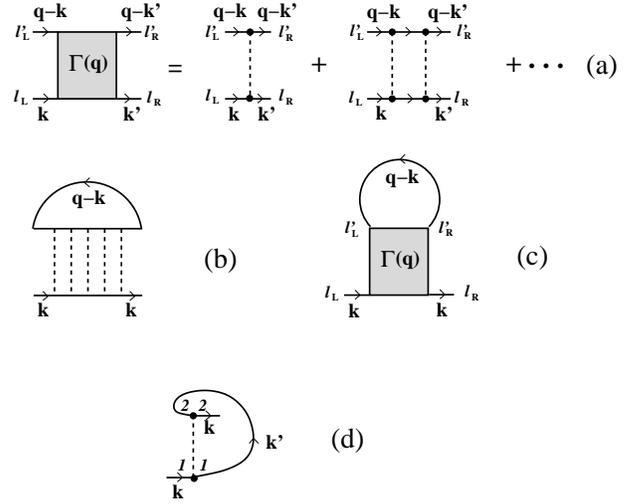}
\vspace{.25truecm}
\caption{
(a) Particle-particle ladder in the broken-symmetry
phase. Conventions for four-momenta and Nambu indices are specified. 
Dots delimiting the potential (broken line) represent $\tau_3$ Pauli 
matrices. 
Only combinations with $\ell_{L}=\ell'_{L}$ and $\ell_{R}=\ell'_{R}$  
occur owing to the regularization we have adopted for the potential.
(b) Fermionic self-energy diagram associated with the
expression (\ref{Sigma-normal-phase}) in the normal phase.
(c) Fermionic self-energy diagram associated with
the expressions (\ref{Sigma-broken-11}) and (\ref{Sigma-broken-12}) in
the broken-symmetry phase.
(d) BCS contribution (15) to the self-energy.}
\end{figure}

Note further that the above result for $v_2(0)$ can be cast in the bosonic 
form 
$v_2(0) = 4\pi a_B/m_B$ with $a_B = 2 a_F$. The present theory thus describes
the effective interaction between the composite bosons within the Born 
approximation, while improved theories\cite{Pi-S-98,petrov} for $a_B$ would
give smaller values for the ratio $a_B/a_F$.
These improvements will not be considered in the present paper.

Apart from the overall factor $- 8 \pi/(m^{2} a_{F})$ (and a sign difference 
in the off-diagonal component~\cite{APS-02}), the expressions
(\ref{Gamma-11-approx}) and
(\ref{Gamma-12-approx}) coincide with the normal and anomalous
non-condensate bosonic Green's functions
within the Bogoliubov approximation~\cite{FW}, respectively.
These expressions will be specifically exploited in Sec.~IID, where the
strong-coupling limit of the fermionic self-energy will be analyzed in detail.

In the normal phase, on the other hand, the BCS single-particle Green's
functions are replaced by the bare single-particle propagator 
${\mathcal G}_0(p) = [i\omega_{n} - \xi({\mathbf p})]^{-1}$,
while for arbitrary coupling the particle-particle ladder acquires  the form:
\begin{eqnarray}
&&\Gamma_0(q) = - \left\{\frac{m}{4 \pi a_{F}} + \int \! 
\frac{d{\mathbf p}}{(2\pi)^{3}}\right. \nonumber\\  
&&\times \left[\frac{\tanh(\beta \xi({\mathbf p})/2)
+\tanh(\beta
\xi({\mathbf p-q})/2)}{2(\xi({\mathbf p})+\xi({\mathbf p-q})-i\Omega_{\nu})}
\left. - \frac{m}{{\mathbf p}^{2}} \right] \right\}^{-1}
\, .          \label{most-general-pp-sc}
\end{eqnarray}
In particular, in the strong-coupling limit the expression
(\ref{most-general-pp-sc}) reduces to
\begin{equation}
\Gamma_0(q) \, \simeq \, - \, \frac{8 \pi}{m^{2} a_{F}} \, \frac{1}{i
\Omega_{\nu} - {\mathbf q}^{2}/(4m)}\, ,
\label{Gamma-o-approx}
\end{equation}
which coincides (apart again from the overall factor $- 8 \pi/(m^{2}
a_{F})$) with the free-boson Green's function.

The above quantities constitute the essential ingredients of our theory for
the fermionic self-energy and related quantities in the broken-symmetry phase.
As shown in Ref.~\onlinecite{APS-02}, they also
serve to establish a \emph{mapping\/} between
the fermionic and bosonic diagrammatic structures in the broken-symmetry phase,
in a similar fashion to what was done in the normal phase \cite{Pi-S-98}.

\subsection{Choice of the self-energy}

In a recent study \cite{PPSC-02} of the
single-particle spectral function in the normal
phase based on the BCS-BEC crossover approach, the fermionic self-energy was
taken of the form:
\begin{equation}
\Sigma_0(k) \, = \, - \, \frac{1}{\beta {\mathcal V}} \, \sum_{q} \,
\Gamma_0(q) \,\, {\mathcal G}_0(q-k)
\label{Sigma-normal-phase}
\end{equation}
where 
${\mathcal V}$ is the quantization volume and $k=({\bf k},\omega_s)$ is again
a four-vector notation with wave vector ${\bf k}$ and fermionic Matsubara 
frequency $\omega_s$ ($s$ integer).
In this expression, $\Gamma_0(q)$ is given by
Eq.~(\ref{most-general-pp-sc}) for arbitrary coupling and
${\mathcal G}_0(k)$ is the bare single-particle 
propagator.
The self-energy diagram corresponding to the expression
(\ref{Sigma-normal-phase}) is depicted in Fig.~1(b).
The fermionic single-particle excitations are effectively coupled to a 
(bosonic)
superconducting fluctuation mode, which
reduces to a free composite boson in the strong-coupling limit.
Physically, the choice (\ref{Sigma-normal-phase}) for the self-energy
entails the presence of a pairing interaction
above $T_{c}$, which can have significant influence on the single-particle (as
well as other) properties.

In the present paper, we choose the self-energy in the broken-symmetry 
phase below $T_c$, with the aim of recovering the expression 
(\ref{Sigma-normal-phase}) when
approaching $T_c$ from below and the Bogoliubov approximation for the 
composite bosons in the strong-coupling limit.
To this end, we adopt the \emph{simplest\/} approximations to describe
fermionic \emph{as well as\/} bosonic excitations
in the broken-symmetry phase, which reduce to bare fermionic and free
bosonic excitations in the normal phase,
respectively.
These are the BCS single-particle Green's functions
(\ref{BCS-Green-function}) (in the place of the bare single-particle
propagator ${\mathcal G}_0$) and the particle-particle ladder
(\ref{Gamma-solution}) (in the place of its
normal-phase counterpart $\Gamma_0$).
By this token, the fermionic self-energy (\ref{Sigma-normal-phase}) is
replaced by the following $2 \times 2$ matrix:
\begin{eqnarray}
\!\!\!\!\!\!\Sigma^{L}_{11}(k) = -\Sigma^{L}_{22}(-k) =  
- \frac{1}{\beta {\mathcal V}} \sum_{q} 
\Gamma_{11}(q)  {\mathcal G}_{11}(q-k)
\label{Sigma-broken-11}\\
\Sigma^{L}_{12}(k) = \Sigma^{L}_{21}(k)   =  
- \frac{1}{\beta {\mathcal V}}  \sum_{q} 
\Gamma_{12}(q)  {\mathcal G}_{12}(q-k)\phantom{111}    
\label{Sigma-broken-12}
\end{eqnarray}
where the label $L$ refers to the particle-particle ladder.
The corresponding self-energy diagram is depicted in 
Fig.~1(c).\cite{footnote-Nambu-arrows}

The choice (\ref{Sigma-broken-11}) and (\ref{Sigma-broken-12}) for the self
energy is made on physical grounds.
A formal ``ab initio'' derivation of these expressions can also be done in
terms of ``conserving approximations'' in
the Baym-Kadanoff sense, that hold even in the broken-symmetry phase
\cite{Baym-62}.
In such a formal derivation, however, the single-particle Green's functions
entering Eqs.(\ref{Sigma-broken-11}) and
(\ref{Sigma-broken-12}) (also through the particle-particle ladder
(\ref{Gamma-solution})) would be required to be self-consistently determined 
with the \emph{same\/} self-energy insertions.
In our approach, we take instead the single-particle Green's
functions to be of the BCS form (\ref{BCS-Green-function}). The order 
parameter $\Delta$ and chemical potential $\mu$ are obtained, however, via two 
coupled equations (to be discussed in Sec.~IIC) that include the 
self-energy insertions (\ref{Sigma-broken-11}) and (\ref{Sigma-broken-12}).
In this way, we will recover the Bogoliubov form
(\ref{Gamma-11-approx}) and (\ref{Gamma-12-approx})
for the particle-particle ladder not only at zero temperature but also at
finite temperatures (and, in particular, close
to the Bose-Einstein transition temperature).

The choice (\ref{Sigma-broken-11}) and (\ref{Sigma-broken-12}) for the self
energy is not exhaustive. In the broken-symmetry phase there, in fact, exists 
an additional self-energy contribution that survives
the regularization (\ref{v0}) of the interaction potential in the limit
$k_{0} \rightarrow \infty$, even though it
does not contain particle-particle rungs\cite{Haussmann-footnote}.
This additional self-energy diagram is the ordinary BCS contribution
depicted in Fig.~1(d), with the associated expression
\begin{equation}
\Sigma^{BCS}_{12}(k) \, = \, \Sigma^{BCS}_{21}(k) \, = \, - \, \Delta \,\,
,                    \label{Sigma-12-BCS}
\end{equation}
while the corresponding (Hartree-Fock) diagonal elements vanish with the
regularization we have adopted.
Relating the expression (\ref{Sigma-12-BCS}) to the diagram of
Fig.~1(d) rests on the validity of the BCS gap equation 
[Eq.~(\ref{BCS-gap_equation}) below], for \emph{arbitrary\/} values of the 
chemical potential.
For this, as well as for an additional reason (cf. Sec.~IID), we
shall consistently consider that equation to hold 
for the order parameter $\Delta$.

The choice (\ref{Sigma-12-BCS}) alone would be appropriate to describe the 
system
in the weak-coupling (BCS) limit, where the superconducting fluctuation 
contributions (\ref{Sigma-broken-11}) and (\ref{Sigma-broken-12}) represent 
only small corrections.
In the intermediate- and strong-coupling regions, on the other hand, both
contributions
(\ref{Sigma-broken-11})-(\ref{Sigma-broken-12}) \emph{and\/}
(\ref{Sigma-12-BCS}) might become equally significant
(depending on the temperature range below $T_{c}$).
We thus consider both contributions
\emph{simultaneously\/} and write the fermionic self-energy in
the matrix form:
\begin{eqnarray}
\left(\begin{array}{cc} \Sigma_{11}(k) & \Sigma_{12}(k) \\ \Sigma_{21}(k)
& \Sigma_{22}(k) \end{array} \right)\phantom{11111111111111111111111111}
\nonumber\\ 
 = \left( \begin{array}{cc} \Sigma^{L}_{11}(k) & \Sigma^{L}_{12}(k) +
\Sigma^{BCS}_{12}(k) \\
\Sigma^{L}_{21}(k) + \Sigma^{BCS}_{21}(k) & \Sigma^{L}_{22}(k) \end{array}
\right)\, .             \label{total-self-energy}
\end{eqnarray}
 In the following, however, we shall neglect
$\Sigma^{L}_{12}$ in comparison to $\Sigma^{BCS}_{12}$.
It will, in fact, be proved in Sec.~IID  
that, in strong coupling, $\Sigma^{L}_{12}$ is subleading with respect to 
both $\Sigma^{BCS}_{12}$ and $\Sigma^{L}_{11}$. Inclusion of $\Sigma^{L}_{12}$
is thus not required to properly recover the Bogoliubov description for 
the composite bosons in the strong-coupling limit.

To summarize, the fermionic single-particle Green's functions are obtained
in terms of the bare single-particle propagator ${\mathcal G}_0(k)$ and of 
the self-energy (\ref{Sigma-broken-11}) and (\ref{Sigma-12-BCS}) via the 
Dyson's equation in matrix form:
\begin{eqnarray}
& &\left( \begin{array}{cc} G_{11}^{-1}(k) & G_{12}^{-1}(k)  \\ G_{21}^{-1}(k)
& G_{22}^{-1}(k) \end{array} \right)
 =  \left( \begin{array}{cc} {\mathcal G}_0(k)^{-1} & 0 \\ 0 & -
{\mathcal G}_0(-k)^{-1} \end{array} \right)  \nonumber \\
&& - 
\left( \begin{array}{cc} \Sigma^{L}_{11}(k) & 
\Sigma^{BCS}_{12}(k) \\
\Sigma^{BCS}_{21}(k) & \Sigma^{L}_{22}(k) \end{array}
\right) \label{Dyson-equation} \, .         
\end{eqnarray}
If only the BCS contribution (\ref{Sigma-12-BCS}) to the self-energy were 
retained, the fermionic single-particle Green's functions $G_{ij}(k)$
($i,j=1,2$) would reduce to the BCS form (\ref{BCS-Green-function}). Upon
including, in addition, the fluctuation contribution
(\ref{Sigma-broken-11}) to the self-energy, 
modified single-particle Green's functions result,
which we are going to study as functions of coupling strength and
temperature.
\subsection{Comparison with the Popov approximation for dilute superfluid 
fermions}

The choice of the self-energy (\ref{Sigma-broken-11}) and (\ref{Sigma-12-BCS})
resembles the approximation for the self-energy introduced by 
Popov\cite{Popov-87} for
superfluid fermions in the dilute limit $k_F |a_F| \ll 1$ (with $a_F < 0$).
There is, however, an important difference between the Popov fermionic 
approximation and our theory. We include in 
Eq.~(\ref{Sigma-broken-11}) the full $\Gamma_{11}$ obtained by the matrix 
inversion of Eq.~(\ref{Gamma-solution}); Popov instead neglects $\chi_{12}$ 
therein and approximate $\Gamma_{11}$ by $1/\chi_{11}$, thus removing the 
feedback of the Bogoliubov-Anderson mode on the diagonal fermionic self-energy
$\Sigma_{11}$. Retaining this mode is essential when dealing with the BCS-BEC 
crossover, to describe the composite bosons in the 
strong-coupling limit by the Bogoliubov approximation, as 
discussed in Sec.~IIA. 
Approaching the weak-coupling limit, on the other hand, the presence of the 
Bogoliubov-Anderson mode becomes 
progressively irrelevant and the self-energies coincide in the two theories.
As a check on this point, we have verified that, in the weak-coupling limit 
and at zero temperature, $\Sigma_{11}$ 
obtained by our theory (using the numerical procedures discussed in Sec.~III) 
reduces to $4\pi a_F n /(2 m)$, which is the expression obtained also with 
the Popov
approximation\cite{Popov-87} in the absence of the Bogoliubov-Anderson mode.

There is another difference between the Popov fermionic approximation and our 
theory as formulated in Sec.~IIA, which concerns the off-diagonal fermionic
self-energy $\Sigma_{12}$. 
Our expression~(\ref{Sigma-12-BCS}) for $\Sigma_{12}$ was obtained from the
diagram of Fig.~1(d), where the single particle line represents the 
off-diagonal BCS Green's function of Eq.~(\ref{BCS-Green-function}) with no 
insertion of the diagonal self-energy $\Sigma_{11}$.
Within the Popov approximation, on the other hand, $\Sigma_{12}$ is defined
formally by the same diagram of Fig.~1(d), but with the single-particle line
being fully self-consistent (and thus including $\Sigma_{11}$). 
Since $\Sigma_{11}$ turns out to approach a constant value $\Sigma_0$ in the 
weak-coupling limit (as discussed above), inclusion of 
$\Sigma_{11}\simeq\Sigma_0$ can be simply made by a shift of the chemical 
potential (such that $\mu \to \mu - \Sigma_0$). This shift affects, however, 
the
value of the gap function $\Delta$ in a non-negligible way even in the extreme
weak-coupling limit. Neglecting this shift, in fact, results in a reduction 
 by a factor $e^{1/3}$ of the BCS asymptotic expression 
$(8\epsilon_F/e^2) \exp[\pi/(2 k_F a_F)]$ for
$\Delta$ (where $\epsilon_F=k_F^2/(2m)$). Inclusion of
the shift $\Sigma_0$ is thus important to recover the BCS value for $\Delta$
in the (extreme) weak-coupling limit.

The need to include the constant shift $\Sigma_0$ on the weak-coupling side
of the crossover was also discussed in Ref.~\onlinecite{PPSC-02} while 
studying the spectral function $A({\bf k},\omega)$  in the normal phase 
with the inclusion of pairing fluctuations.
In that context, inclusion of the shift $\Sigma_0$ proved necessary to have 
the pseudogap
depression of $A({\bf k},\omega)$ centered about $\omega=0$.
Inclusion of the shift $\Sigma_0$ in the broken-symmetry phase (at least 
when approaching the critical temperature from below) is thus also necessary
to connect the spectral function $A({\bf k},\omega)$  with continuity 
in the weak-coupling side of the crossover.

Combining the above needs for $\Delta$ and $A({\bf k},\omega)$, we have 
introduced the constant shift $\Sigma_0$ for all temperatures below $T_c$,
by replacing $\mu$ with $\mu-\Sigma_0$ 
in the BCS Green's functions~(\ref{BCS-Green-function}) entering the 
convolutions (\ref{A-definition}) and (\ref{B-definition}). 
The same replacement is made in the gap equation 
[Eq.~(\ref{BCS-gap_equation}) below].
In the Dyson's equation (\ref{Dyson-equation}), however, $\mu$ is left 
unchanged since the constant shift $\Sigma_0$ is already contained in 
$\Sigma_{11}(k)$ as soon as its $k$-dependence is irrelevant. Accordingly, we
have included this constant shift in the calculation of both thermodynamic 
and dynamical quantities in the weak-coupling side for
$(k_F a_F)^{-1}\le -0.5 $, and neglected it for larger couplings 
when $\Sigma_{11}(k)$ can no longer be approximated by a constant. 

\begin{figure}
\includegraphics[scale=0.7]{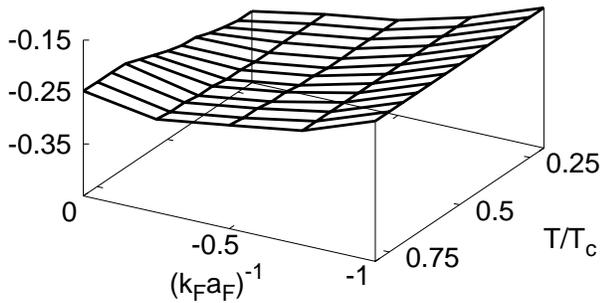}
\vspace{.25truecm}
\caption{Self-energy shift 
$\Sigma_0$ (in units of 
$\epsilon_F$) vs temperature $T$ (units of $T_c$) and 
coupling $(k_F a_F)^{-1}$.}
\label{shift}
\end{figure}

It turns out that the temperature dependence of $\Sigma_0$ is rather weak  
in the above coupling range. A plot of 
$\Sigma_0$ vs $T/T_c$ and $(k_F a_F)^{-1}$ is shown in Fig.~\ref{shift}. 
Here, the critical temperature $T_c$ is obtained by 
applying the Thouless criterion from the normal phase as was done in 
Ref.~\onlinecite{PPSC-02} (this procedure to obtain $T_c$ will be used 
in the rest of the paper). In this plot, the constant shift $\Sigma_0$ is 
obtained as 
$\Sigma_0={\rm Re} \Sigma_{11}^R(|{\bf k}|=
\sqrt{ 2 m (\mu - \Sigma_0)},\omega=0)$, in analogy to what was also done in 
Ref.~\onlinecite{PPSC-02}. Here, $\Sigma_{11}^R({\bf k},\omega)$ 
is the analytic continuation to the real frequency axis of the Matsubara 
self-energy $\Sigma_{11}({\bf k},\omega_s)$ discussed in Sec.~IIE.  

\subsection{Coupled equations for the order parameter and the chemical 
potential}

Thermodynamic quantities, such as the order parameter $\Delta$ and the
chemical potential $\mu$, are obtained directly
in terms of the Matsubara single-particle Green's functions, without
the need of resorting to the analytic continuation to the
real frequency axis.

Quite generally, the order parameter $\Delta$ is defined in terms of the
``anomalous'' Green's function
$G_{12}({\mathbf k},\omega_{s})$ via 
$\Delta=v_{0}\langle\psi_{\uparrow}({\mathbf r})\psi_{\downarrow}({\mathbf r})
\rangle$
[cf. Eq.~(\ref{equation-of-motion})], where the strength
$v_{0}$ of the contact potential is kept to
comply with a standard definition of BCS theory \cite{FW}.
One obtains:
\begin{equation}
\Delta \, = \, - \, v_{0} \, \int \! \frac{d {\mathbf k}}{(2\pi)^{3}} \,
\frac{1}{\beta} \sum_{s} 
            \, G_{12}({\mathbf k},\omega_{s})  \,\, .
\label{Delta-G-12}
\end{equation}
By the same token, the chemical potential $\mu$ can be obtained in terms of 
the ``normal'' Green's function
$G_{11}({\mathbf k},\omega_{s})$ via the particle density 
$n$:
\begin{equation}
n \, = \, 2 \, \int \! \frac{d {\mathbf k}}{(2\pi)^{3}} \, \frac{1}{\beta}
\sum_{s} \, e^{i\omega_{s}\eta}
            \, G_{11}({\mathbf k},\omega_{s})  
\label{n-G-11}
\end{equation}
where $\eta=0^+$.
The two equations (\ref{Delta-G-12}) and (\ref{n-G-11}) are 
coupled, since the Green's functions depend on
both $\Delta$ and $\mu$.
The results of their numerical solution will be presented in the next
section for various temperatures and couplings.

In the following treatment, we shall deal with the two equations
(\ref{Delta-G-12}) and (\ref{n-G-11})
\emph{on a different footing\/}.
Specifically, we will enter in the density equation (\ref{n-G-11})
the expression for the normal Green's function obtained from 
Eq.~(\ref{Dyson-equation}), that includes both BCS \emph{and\/} 
fluctuation contributions (see Eq.~(\ref{G-11-Matsubara}) 
below). We will use instead in the gap equation
(\ref{Delta-G-12}) the BCS anomalous function (\ref{BCS-Green-function}), 
that includes only the BCS self-energy (\ref{Sigma-12-BCS}).
In this way, the gap equation (\ref{Delta-G-12}) reduces to the form
\begin{equation}
\frac{m}{4 \pi a_{F}} \, + \, \int \! \frac{d{\mathbf k}}{(2\pi)^{3}} \, \left[
\frac{\tanh(\beta E({\mathbf k})/2)}{2 E({\mathbf k})} \, - \,
\frac{m}{{\mathbf k}^{2}} \right] \, = \, 0
\label{BCS-gap_equation}
\end{equation}
where the regularization of the contact potential in terms of the
scattering length $a_{F}$ has been introduced.
This equation has the same \emph{formal\/} structure of the BCS
gap equation, although the numerical values
of the chemical potential entering Eq.~(\ref{BCS-gap_equation}) 
differ from those
obtained by the BCS density equation.
This procedure ensures that the bosonic propagators 
(\ref{Gamma-solution})
in the broken-symmetry phase are \emph{gapless\/}, as shown explicitly by the 
Bogoliubov-type expressions
(\ref{Gamma-11-approx}) and (\ref{Gamma-12-approx})
in the strong-coupling limit.
In general, in fact, there is no \emph{a priori\/} guarantee that a given
(conserving) approximation for fermions would result into a ``gapless'' 
approximation~\cite{HM-65}  for the composite bosons in the strong-coupling
limit of the fermionic attraction.

Including fluctuation corrections to the BCS density equation as in 
Eq.~(\ref{n-G-11}), on the other 
hand, results in the emergence of important effects in the strong-coupling 
limit of the theory, as discussed next.

\subsection{Analytic results in the strong-coupling limit}

We proceed to show that the original
fermionic theory, as defined by the Dyson's equation~(\ref{Dyson-equation}), 
maps onto the Bogoliubov theory for the composite bosons which 
form as bound-fermion pairs in the strong-coupling limit.
To this end, we shall exploit the conditions $\beta|\mu|\gg 1$
and $\Delta\ll|\mu|$ ($\mu<0$) (which
\emph{define\/} the strong-coupling limit) in the (Matsubara) expressions
(\ref{Sigma-broken-11}) and (\ref{Sigma-broken-12})
for $\Sigma^{L}(p)$, thus also verifying that $\Sigma_{12}^L$ can be neglected.

 These expressions are calculated by performing the
wave-vector and frequency convolutions with
the approximate expressions (\ref{Gamma-11-approx}) and
(\ref{Gamma-12-approx}) for the particle-particle ladder
and the expressions (\ref{BCS-Green-function}) for the BCS single-particle
Green's functions.

 Upon neglecting contributions that are subleading under the
above conditions, we obtain in this way for the diagonal part of the 
self-energy:
\begin{eqnarray}
\Sigma_{11}^{L}({\mathbf k},\omega_{s}) & \simeq & \frac{8 \pi}{m^{2} a_{F}} 
\int \! \frac{d {\mathbf q}}{(2 \pi)^{3}} 
\left[   \frac{u_{B}^{2}({\mathbf q}) 
b(E_{B}({\mathbf q}))}
{i\omega_{s} + E({\mathbf q}-{\bf k}) - E_{B}({\mathbf q})} \right.\nonumber \\
& - &  \left.  \frac{v_{B}^{2}({\mathbf q}) 
b(-E_{B}({\mathbf q}))}
{i\omega_{s} + E({\mathbf q}-{\bf k}) + E_{B}({\mathbf q})}   \right]   \, .
\label{Sigma-11-strong-coupling}
\end{eqnarray}
In this expression, $E({\mathbf k})$ is the BCS dispersion of 
Eqs.~(\ref{BCS-Green-function}),
$E_{B}({\mathbf q})$ is the Bogoliubov dispersion
relation (\ref{Bogoliubov-disp}), $b(x) = [\exp (\beta x) - 1]^{-1}$ is
the Bose distribution, and
\begin{equation}
v_{B}^{2}({\mathbf q}) \, = \, u_{B}^{2}({\mathbf q}) - 1 \, = \,
\frac{\frac{{\mathbf q}^{2}}{2m_{B}} \, + \mu_{B} \, - \,
E_{B}({\mathbf q})}{2 E_{B}({\mathbf q})}   \label{u-v-Bogoliubov}
\end{equation}
are the standard bosonic factors of the Bogoliubov transformation \cite{FW}.

In the numerators of the expressions within brackets in 
Eq.~(\ref{Sigma-11-strong-coupling}), the Bose functions are peaked at about
${\mathbf q}=0$ and vary over a scale
${\mathbf q}^{2}/(2m_{B}) \approx T \ll |\mu|$. Similarly, the factors
$u_{B}^{2}({\mathbf q})$ and
$v_{B}^{2}({\mathbf q})$ are also peaked at about
${\mathbf q}=0$ and vary over a scale
${\mathbf q}^{2}/(2m_{B}) \approx \mu_{B} \ll |\mu|$.
The denominators in the expression (\ref{Sigma-11-strong-coupling}), on the
other hand, vary over the much larger
scale $|\mu|$. 
%[Note that the same approximations if $i\omega_s$ would be
%replaced by a real number. This point will be discussed in more detail in 
%Secs. IIE and IIIB.]
For these reasons, we can further approximate the expression
(\ref{Sigma-11-strong-coupling}) as follows:
\begin{equation}
\Sigma_{11}^{L}({\mathbf k},\omega_{s}) \, \simeq \, \frac{8 \pi}{m^{2}
a_{F}} \,\,
\frac{1}{i\omega_{s} + \xi({\mathbf k})} \,\, n'_{B}(T)
\label{Sigma-11-n-prime}
\end{equation}
where
\begin{equation}
n'_{B}(T) = 
\int \! \frac{d {\mathbf q}}{(2 \pi)^{3}} \left[
u_{B}^{2}({\mathbf q}) b(E_{B}({\mathbf q}))
- v_{B}^{2}({\mathbf q}) b(-E_{B}({\mathbf q})) \right]
\label{noncondensate-density}
\end{equation}
identifies the bosonic \emph{noncondensate density\/} according to
Bogoliubov theory \cite{FW}.
Note that in the normal phase (when the condensate density $n_0(T)$ of
Eq.~(\ref{pot-chim-Bog}) vanishes),
the noncondensate density (\ref{noncondensate-density}) becomes the full
bosonic density $n_{B}=n/2$, and Eq.~(\ref{Sigma-11-n-prime}) reduces to the 
expression obtained in Ref.~\onlinecite{Pi-S-98} directly from the form
(\ref{Sigma-normal-phase}) of the fermionic self-energy.

The off-diagonal self-energy $\Sigma_{12}^{L}(k)$ can be analyzed in a
similar way.
Since its magnitude is supposed to be the largest at zero temperature, we
estimate it correspondingly for
${\mathbf k}=0$ and $\omega_{s}=0$ as follows:
\begin{eqnarray}
\Sigma_{12}^{L}({\mathbf k}=0,\omega_{s}=0) &\simeq& \frac{8 \pi}{m^{2}
a_{F}} \frac{\mu_{B} \Delta }{2} 
\int \! \frac{d {\mathbf k'}}{(2 \pi)^{3}} \frac{1}{E({\mathbf k'}) 
E_{B}({\mathbf k'})}\nonumber\\
&\times& \frac{1}{(E({\mathbf k'}) \, +\,  E_{B}({\mathbf k'}))}  \,\, .
\label{Sigma-12-strong-coupling}
\end{eqnarray}
At the leading order, we can neglect both $\Delta$ and $\mu_{B}$ in the
integrand, where the energy scale $|\mu|$
dominates.
We thus obtain $\Sigma_{12}^{L}(k=0) \approx \Delta (\Delta^{2}/(2
|\mu|^{2}))$ in the strong-coupling limit (where the relation 
$\mu_{B}=\Delta^{2}/(4|\mu|)$ - see below - has been used). This represents a 
subleading contribution
in the small dimensionless parameter $\Delta/|\mu|$ with respect to both
the BCS contribution $\Sigma_{12}^{{\rm BCS}}(k)=-\Delta$ 
and the diagonal fluctuation contribution $\Sigma_{11}^{L}(k)$. It 
can accordingly be neglected.

Within the above approximations, the inverse~(\ref{Dyson-equation}) of the 
fermionic single-particle Green's 
function reduces to:
\begin{eqnarray}
\left( \begin{array}{cc} G_{11}^{-1}(k) & G_{12}^{-1}(k)  \\ G_{21}^{-1}(k)
& G_{22}^{-1}(k) \end{array} \right)
\phantom{1111111111111111111111111}\nonumber\\
\simeq \left( \begin{array}{cc} i\omega_{s}-\xi({\mathbf k}) -
\frac{\Delta_0^{2}}{i\omega_{s}+\xi({\mathbf k})} &  \Delta \\
 \Delta & i\omega_{s}+\xi({\mathbf k}) -
\frac{\Delta_0^{2}}{i\omega_{s}-\xi({\mathbf k})} \end{array} \right)
\label{G-inverse-strong-coupling}
\end{eqnarray}
with the notation
\begin{equation}
\Delta_0^{2} \, \equiv \, \frac{8 \pi}{m^{2} a_{F}} \,\, n'_{B}(T)  \;
.   \label{Delta-o}
\end{equation}
From Eq.~(\ref{G-inverse-strong-coupling}) we get the desired
expression for $G_{11}({\mathbf k},\omega_{s})$
in the strong-coupling limit:
\begin{equation}
G_{11}({\mathbf k},\omega_{s}) \, \simeq \, \frac{1}{i\omega_{s} \, - \,
\xi({\mathbf k}) \, - \,
\frac{\Delta^{2} \, + \, \Delta_0^{2}}{i\omega_{s} \, + \,
\xi({\mathbf k})}}                 \label{G-11-strong-coupling}
\end{equation}
where we have discarded a term of order $\Delta_0^{2}/|\mu|$ with respect
to $|\mu|$.
Note that Eq.~(\ref{G-11-strong-coupling}) has the same formal structure of 
the corresponding BCS expression (\ref{BCS-Green-function}), with the 
replacement 
$E({\mathbf k})
\rightarrow
\tilde{E}({\mathbf k})=\sqrt{\xi({\mathbf k})^{2}+(\Delta^{2}+\Delta_0^{2})}
$.
We rewrite it accordingly as:
\begin{equation}
G_{11}({\mathbf k},\omega_{s}) \, \simeq \,
\frac{\tilde{u}^{2}({\mathbf k})}{i\omega_{s} \, - \, \tilde{E}({\mathbf k})}
\, + \, \frac{\tilde{v}^{2}({\mathbf k})}{i\omega_{s} \, + \,
\tilde{E}({\mathbf k})}        
\label{G-11-strong-coupling-figo}
\end{equation}
with the modified BCS coherence factors $\tilde{v}^{2}({\mathbf k}) = 1 -
\tilde{u}^{2}({\mathbf k}) =
(1 - \xi({\mathbf k})/\tilde{E}({\mathbf k}))/2$.

Before making use of the asymptotic expression
(\ref{G-11-strong-coupling-figo}) in the density equation (\ref{n-G-11}),
it is convenient to manipulate suitably the gap equation
(\ref{BCS-gap_equation}) in the strong-coupling limit.
Expanding $1/E({\mathbf k})$ therein as
$[1-\Delta^{2}/(2\xi({\mathbf k})]/\xi({\mathbf k})$  and evaluating the
resulting elementary integrals, one obtains:
\begin{equation}
\frac{\Delta^{2}}{4|\mu|} \, \simeq \, 2 \, \left( \sqrt{2 \, |\mu| \,
\epsilon_0} \, - \, 2 \, |\mu| \right)\, .
\label{Delta-mu-strong-coupling}
\end{equation}
Setting further $2 \mu = - \epsilon_0 + \mu_{B}$, one gets the
relation $\Delta^{2}/(4|\mu|)=\mu_{B}$ quoted already 
after Eqs.~(\ref{Bogoliubov-disp}) and~(\ref{Sigma-12-strong-coupling}).

Let's now consider the density equation (\ref{n-G-11}).
With the BCS-like form (\ref{G-11-strong-coupling-figo}) one obtains
immediately:
\begin{equation}
n \, \simeq \, 2 \, \int \! \frac{d {\mathbf k}}{(2\pi)^{3}} \,
\tilde{v}^{2}({\mathbf k})        \label{n-G-11-strong-coupling}
\end{equation}
that holds for $T \ll \epsilon_{0}$, at temperatures well below the
dissociation threshold of the composite
bosons.
Similarly to what was done to get the gap equation 
(\ref{Delta-mu-strong-coupling}), in Eq.~(\ref{n-G-11-strong-coupling}) one  
expands $1/\tilde{E}({\mathbf k})$ as
$[1-(\Delta^{2}+\Delta_0^{2})/(2\xi({\mathbf k})]/\xi({\mathbf k})$  and
evaluates the resulting elementary integrals, to obtain:
\begin{equation}
n \, \simeq \, \frac{m^{2} \, a_{F}}{4 \pi} \,\, \left( \Delta^{2} \, + \,
\Delta_0^{2} \right) \,\, .  \label{final-n}
\end{equation}
Recalling the definition (\ref{Delta-o}) for $\Delta_{0}^{2}$,
as well as the expressions
(\ref{Delta-mu-strong-coupling}) and (\ref{pot-chim-Bog}) for the order
parameter, which we rewrite in the form
\begin{equation}
\Delta^{2} \, = \, \frac{8 \pi}{m^{2} a_{F}} \,\, n_0(T)
\label{Delta-n-o}
\end{equation}
in analogy to Eq.~(\ref{Delta-o}), the result (\ref{final-n}) 
becomes eventually:
\begin{equation}
n \, = \, 2 \left( n'_{B}(T) \, + \, n_0(T) \right)
\label{n-n-B-n-o}
\end{equation}
that holds asymptotically for $T \ll \epsilon_0$.

These results imply that, in the strong-coupling limit, the original fermionic
theory recovers the Bogoliubov theory for the
composite bosons, not only at zero temperature but also \emph{at any
temperature\/} in the broken-symmetry phase.
Accordingly, the noncondensate density $n'_{B}(T)$ is given by the
expression (\ref{noncondensate-density}), the bosonic
factors $v_{B}^{2}({\mathbf q})$ and $u_{B}^{2}({\mathbf q})$ are given by
Eq.~(\ref{u-v-Bogoliubov}), and the dispersion
relation $E_{B}({\mathbf q})$ is given by Eq.~(\ref{Bogoliubov-disp}).
In the strong-coupling limit, the present fermionic theory thus inherits all 
virtues and shortcomings of the Bogoliubov theory for a weakly-interacting 
Bose gas~\cite{Bassani-GCS-01}.
The present fermionic theory at arbitrary coupling then provides 
an interpolation procedure 
between the Bogoliubov theory for the composite bosons and the weak-coupling 
BCS theory plus pairing fluctuations.
Both these analytic limits will constitute important checks on the
numerical calculations reported in Sec.~III.
Note that inclusion of the off-diagonal fluctuation contribution 
$\Sigma_{12}^L(k)$ to the 
self-energy is not required to recover the Bogoliubov theory in strong 
coupling. For this reason, we will not consider $\Sigma_{12}^L(k)$ altogether
in the numerical calculations presented in Sec.~III, as anticipated in 
Eq.~(\ref{Dyson-equation}).

The above analytic results enable us to infer the main features of the
temperature dependence of the order parameter in the strong-coupling limit.
In particular, the low-temperature behavior 
$n_0(T) = n_0(0) - m_{B}(k_{B} T)^{2}/(12 c)$( where
$c = \sqrt{n_0v_{2}(0)/m_{B}}$ is the sound  velocity) within the
Bogoliubov approximation, implies that
$\Delta(T)$ decreases from $\Delta(0)$ with a $T^{2}$ behavior, in
the place of the exponential behavior
obtained within the BCS theory (with an $s$-wave order parameter) \cite{FW}.
In addition, in the present theory the order parameter vanishes over the
scale of the Bose-Einstein transition temperature $T_{BE}$, while in the BCS 
theory it would vanish over the scale of the bound-state energy
$\epsilon_0$ of the composite bosons.

Note finally that the fermionic quasi-particle dispersion
$\tilde{E}({\mathbf k})$, entering the expression
(\ref{G-11-strong-coupling-figo}) of the diagonal Green's function in the
strong-coupling limit, contains the sum $\Delta^{2}+\Delta_0^{2}$
instead of the single term $\Delta^{2}$ of the BCS dispersion
$E({\mathbf k})$.

\subsection{Spectral function and sum rules}

We pass now to identify the form of the spectral function
$A({\mathbf k},\omega)$ associated with the approximate choice 
 of the Matsubara self-energy of Eq.~(\ref{Dyson-equation}).
To this end, we need to perform the \emph{analytic continuation\/} in the
complex frequency plane, thus determining
the {\em retarded\/} fermionic single-particle Green's functions from their
Matsubara counterparts.
The approach developed in this subsection holds specifically for the 
approximate choice for the self-energy of Eq.~(\ref{Dyson-equation}). It thus 
differs from the general analysis presented in the Appendix which holds
for the exact Green's functions, irrespective of any specific approximation.

In general, the process of analytic continuation to the real frequency axis 
from the numerical Matsubara Green's functions proves
altogether nontrivial, as it requires in practice recourse to approximate 
numerical methods such as, e.g., the method of Pad\'{e} 
approximants \cite{Serene-1977}.
We then prefer to rely on a procedure whereby the analytic 
continuation to the real frequency axis is achieved by avoiding numerical 
extrapolations from the Matsubara Green's functions.

The fermionic normal and anomalous Matsubara single-particle
Green's functions are obtained at any given coupling from matrix inversion of 
Eq.~(\ref{Dyson-equation}):
\begin{eqnarray}
&&G_{11}({\mathbf k},\omega_{s}) = \left[\phantom{\frac{1}{1}}\!\!\!
i\omega_{s}-\xi({\mathbf k})-\Sigma_{11}({\mathbf k},\omega_{s})\right.
\nonumber\\
&&\phantom{111111111}-\left.
\frac{\Delta^2}{i\omega_{s}  +  \xi({\mathbf k})
- \Sigma_{22}({\mathbf k},\omega_{s})}\right]^{-1}
\label{G-11-Matsubara}  \\
&&G_{12}({\mathbf k},\omega_{s})= 
\Delta [\,(i\omega_{s}-\xi({\mathbf k})-
\Sigma_{11}({\mathbf k},\omega_{s}))\nonumber\\
&&\phantom{1111} \times  (i\omega_{s}+\xi({\mathbf k})-
\Sigma_{22}({\mathbf k},\omega_{s})) - \Delta^{2}]^{-1}\; . 
\label{G-12-Matsubara} 
\end{eqnarray} 

Consider first the normal Green's function~(\ref{G-11-Matsubara}), which we rewrite in the compact form
\begin{equation}
G_{11}({\mathbf k},\omega_{s}) \, = \,
\frac{1}{i\omega_{s}-\xi({\mathbf k})-\sigma_{11}({\mathbf k},\omega_{s})}
\label{G-11-compact}
\end{equation}
with the short-hand notation
\begin{equation}
\sigma_{11}({\mathbf k},\omega_{s}) \, \equiv \,
\Sigma_{11}({\mathbf k},\omega_{s}) \, + \,
\frac{\Delta^2}{i\omega_{s} \,\, + \,\, \xi({\mathbf k})
- \,\, \Sigma_{22}({\mathbf k},\omega_{s})} \,\, .
\label{sigma-11-compact}
\end{equation}
To perform the analytic continuation of this expression, we look for a 
function $\sigma_{11}({\mathbf k},z)$ of the complex frequency $z$ which 
satisfies the following {\em requirements\/} at any given ${\mathbf k}$:\\
(i) It is analytic off the real axis;\\
(ii) It reduces to $\sigma_{11}({\mathbf k},\omega_{s})$ given by 
Eq.~(\ref{sigma-11-compact}) when $z$ takes the discrete values 
$i \omega_{s}$ on the imaginary axis;\\
(iii) Its imaginary part is negative (positive) for $\mathrm{Im} \, z>0$
($\mathrm{Im} \, z<0$) ;\\
(iv) It vanishes when $|z|\to\infty$ along any straight line parallel to the
real axis with $\mathrm{Im} \, z \neq 0$.

Once the function $\sigma_{11}({\mathbf k},z)$ is obtained,
the expression
\begin{equation}
G^{R}({\mathbf k},\omega) \, = \, \frac{1}{\omega \, + \, i\eta \, - \,
\xi({\mathbf k}) \, - \,
\sigma_{11}({\mathbf k},\omega + i\eta)}
\label{G-R-compact}
\end{equation}
($\eta$ being a positive infinitesimal) represents
the \emph{retarded\/} ($R$) single-particle Green's
function (for real $\omega$) associated with the Matsubara Green's function
(\ref{G-11-compact}), since it satisfies the requirements of the
Baym-Mermin theorem~\cite{Baym-Mermin-61} for the analytic continuation 
from the Matsubara Green's function.

The first step of the above program is to find the analytic continuation of 
$\Sigma_{11}({\mathbf k},\omega_s)$ (and  
$\Sigma_{22}({\mathbf k},\omega_s)$) off the real axis in the complex 
$z$-plane. To this end, it is convenient to express 
$\Sigma_{11}({\mathbf k},\omega_s)$ via the spectral form:
\begin{equation}
\Sigma_{11}({\mathbf k},\omega_s)=
\int_{-\infty}^{+\infty}\frac{d\omega'}{\pi}
\frac{h({\mathbf k},\omega')}{i\omega_s-\omega'}
\label{spect}\;\, . 
\end{equation}
With the replacement $i \omega_s\to z$, the spectral representation 
(\ref{spect}) defines an analytic function $\Sigma_{11}({\mathbf k},z)$ off 
the real axis. In the case of interest with $\Sigma_{11}({\mathbf k},\omega_s)$
given by Eq.~(\ref{Sigma-broken-11}), the function $h({\mathbf k},\omega)$
of Eq.~(\ref{spect}) reads:
\begin{eqnarray}
h({\bf k},\omega)&=&-\int \frac{d {\bf q}}{(2\pi)^3}
\left\{u^2_{{\bf q}-{\bf k}} 
{\rm Im}\,\Gamma_{11}^R({\bf q},\omega+E({\bf q}-{\bf k}))\right.\nonumber\\ 
&\times& \left[f(E({\bf q}-{\bf k}))+b(\omega+E({\bf q}-{\bf k}))\right] 
\nonumber 
\\
&+ & v^2_{{\bf q}-{\bf k}} {\rm Im}\,\Gamma_{11}^R({\bf q},
\omega-E({\bf q}-
{\bf k}))\nonumber\\
 &\times& \left.\left[f(-E({\bf q}-{\bf k}))+b(\omega-E({\bf q}-{\bf k}))
\right] \right\} 
\label{imsig11}
\end{eqnarray}
where $f(x)=[\exp(\beta x)+1]^{-1}$ is the Fermi distribution while 
$u^2_{{\bf k}}$ and $v^2_{{\bf k}}$ are the BCS coherence factors.
To obtain the expression (\ref{imsig11}), a spectral representation has been 
also introduced for $\Gamma_{11}$ entering Eq.~(\ref{Sigma-broken-11}), by 
writing:
\begin{equation}
\Gamma_{11}({\mathbf q},\Omega_{\nu}) \, = - \frac{1}{\pi} 
\int_{-\infty}^{+\infty} \! d\omega' \,
\frac{{\rm Im}\, \Gamma_{11}^R({\mathbf q},\omega')}{i\Omega_{\nu} - 
\omega'}\; .
\label{spectral-representation}
\end{equation}
Here, the spectral function $\Gamma_{11}^R({\mathbf q},\omega)$ is
\emph{defined\/} by  
$\Gamma_{11}({\mathbf q},i\Omega_{\nu} \rightarrow \omega + i \eta)$, 
which is obtained
from the definitions (\ref{Gamma-solution})-(\ref{B-definition}) with the
replacement
$i\Omega_{\nu} \rightarrow \omega + i \eta$ {\em after} the sum over the
internal frequency $\omega_n$ has been performed therein. Even in the absence 
of an explicit Lehmann representation for $\Gamma_{11}$, in fact, it can 
be shown that the spectral representation (\ref{spectral-representation}) 
holds provided the 
function  $\Gamma_{11}({\mathbf q},i\Omega_{\nu}\to z)$ of the complex variable
$z$ is analytic off the real axis. The crucial point is to verify
that the denominator in Eq.~(\ref{Gamma-solution}) with the replacement 
$i \Omega_{\nu}\to z$ never vanishes off the real axis. This property can be
explicitly verified in the strong-coupling limit, as discussed below. For 
arbitrary coupling, we have checked it with the help of numerical calculations.
For the validity of the expression $(\ref{spectral-representation})$, it is 
also required that $\Gamma_{11}({\mathbf q},z)$ vanishes for $|z|\to\infty$. 
This property can be proved directly from 
Eqs.~(\ref{Gamma-solution})-(\ref{B-definition}), according to which  
$\Gamma_{11}({\mathbf q},z)$ has the asymptotic expression
\begin{equation}
\Gamma_{11}({\bf q},z) \simeq \frac{-1}{\frac{m}{4 \pi a_{F}}- 
\frac{m^{3/2}}{4 \pi} \sqrt{-z +{q^2 \over 4 m} - 
2 \mu}}
\label{gammasy}
\end{equation}
and thus vanishes for $|z|\to \infty$. 
Once $\Sigma_{11}({\mathbf k},z)$ has been explicitly constructed according to 
the above prescriptions, $\Sigma_{22}({\mathbf k},z)$ is obtained as 
$-\Sigma_{11}({\mathbf k},-z)$ in accordance with Eq.~(\ref{Sigma-broken-11}).

From the spectral representation (\ref{spect}) for 
$\Sigma_{11}({\mathbf k},z)$, it can be further shown that 
$\Sigma_{11}({\mathbf k},z)$  vanishes when $|z|\to\infty$ along any straight
line parallel to the real axis with ${\rm Im}\,\, z \neq 0$. It can also be shown
that ${\rm Im}\,\Sigma_{11}({\mathbf k},z) < 0$ 
(${\rm Im}\,\Sigma_{11}({\mathbf k},z) > 0$) when ${\rm Im}\, z > 0$ 
(${\rm Im}\, z < 0$). This property follows from the spectral representation of
 $\Sigma_{11}({\mathbf k},z)$, provided $h({\bf k},\omega)\ge 0$ in 
Eq.~(\ref{spect}). For arbitrary coupling, we have verified that
$h({\bf k},\omega)\ge 0$ with the help of numerical calculations. In the 
strong-coupling limit, this condition can be explicitly proved, as discussed 
below.

From these properties of $\Sigma_{11}({\mathbf k},z)$  
(and  $\Sigma_{22}({\mathbf k},z)$) it can then be verified that 
the function
\begin{eqnarray}
\sigma_{11}({\mathbf k},z) & = &
\Sigma_{11}({\mathbf k},z) \nonumber\\
&+&
\frac{\Delta^{2}}{z \, + \, \xi({\mathbf k})
+ \, \Sigma_{11}({\mathbf k},-z)} ,
\label{sigma-tilde-compact}
\end{eqnarray}
satisfies the requirements (i)-(iv) stated after Eq.~(\ref{sigma-11-compact}).
With the replacement $z \to \omega + i\eta$, Eq.~(\ref{G-R-compact}) follows 
eventually on the real frequency axis for the retarded Green's function 
$G^R({\bf k},\omega)$. 

For later convenience, we introduce the following notation on the real 
frequency axis :
\begin{equation}
\Sigma_{11}^R({\mathbf k},\omega) \equiv 
\Sigma_{11}({\mathbf k},\omega + i\eta)
\end{equation}  
such that $\Sigma_{11}({\mathbf k},-\omega - i\eta) = 
\Sigma_{11}^R({\mathbf k},-\omega)^*$ and 
\begin{eqnarray}
\sigma_{11}^R({\mathbf k},\omega)\equiv 
\sigma_{11}({\mathbf k},\omega + i\eta)\phantom{1111111111111111111}\nonumber\\
=\Sigma_{11}^R({\mathbf k},\omega) \, + \,
\frac{\Delta^{2}}{\omega + i \eta \, + \, \xi({\mathbf k})
+ \, \Sigma_{11}^R(-{\mathbf k},-\omega)^*}\; .
\label{sigma-tilde-real}
\end{eqnarray}
From Eq.~(\ref{spect}) it is also clear that 
${\rm Im}\,\Sigma_{11}^R({\mathbf k},\omega)= -h({\mathbf k},\omega)$, and that
${\rm Re}\Sigma_{11}^R({\mathbf k},\omega)$ and 
${\rm Im}\,\Sigma_{11}^R({\mathbf k},\omega)$ are related by a Kramers-Kronig 
transform.

As anticipated, the properties of the function $\Sigma_{11}({\bf k},z)$, 
required above to obtain 
the retarded Green's function (\ref{G-R-compact}) on the real axis, can
be explicitly verified in the strong-coupling limit without recourse to 
numerical calculations. In this case, the approximate expression 
(\ref{Gamma-11-approx}) can be used for $\Gamma_{11}$.  This can be cast in 
the form (\ref{spectral-representation}), with
\begin{eqnarray}
{\rm Im}\, \, \Gamma_{11}^R({\bf q},\omega)&=& -\frac{8 \pi^2}{m^{2}
a_{F}}  
[v_{B}^2({\bf q})\delta(\omega + E_{B}({\mathbf q}))\nonumber\\
&&\phantom{11111}- u_{B}^2({\bf q})\delta(\omega - E_{B}({\mathbf q}))]\; .
\label{imgamma}
\end{eqnarray} 
Entering the expression (\ref{imgamma}) into Eq.~(\ref{imsig11}) and the 
resulting expression into Eq.~(\ref{spect}), one obtains for
$\Sigma_{11}({\bf k},\omega_s)$ the sum of four terms:
\begin{eqnarray}
&&\Sigma_{11}({\bf k},\omega_s)=-\frac{8 \pi}{m^{2} a_{F}}\nonumber\\
&&\times 
\int \! \frac{d {\mathbf q}}{(2 \pi)^{3}} 
\left\{   u_{B}^{2}({\mathbf q}) u^2_{{\bf q}- {\bf k}}
\frac{b(E_{B}({\mathbf q})) + f(E({\bf q}-{\bf k}))}{E_B({\bf q}) - 
E({\bf q} - {\bf k}) - i\omega_{s}}   \right.                   \nonumber \\
& + &  \left. u_{B}^{2}({\mathbf q}) v^2_{{\bf q}- {\bf k}}
\frac{b(E_{B}({\mathbf q})) + f(- E({\bf q}-{\bf k}))}{E_B({\bf q}) + 
E({\bf q} - {\bf k}) - i\omega_{s}}   \right. \nonumber\\
& + & \left.  
v_{B}^{2}({\mathbf q})u^2_{{\bf q}- {\bf k}}
\frac{b(-E_{B}({\mathbf q})) + f(E({\bf q}-{\bf k}))}{E_B({\bf q}) + 
E({\bf q} - {\bf k}) + i\omega_{s}}   \right. \nonumber\\  
& + & \left.  
v_{B}^{2}({\mathbf q})v^2_{{\bf q}- {\bf k}}
\frac{b(-E_{B}({\mathbf q})) + f(-E({\bf q}-{\bf k}))}{E_B({\bf q}) - 
E({\bf q} - {\bf k}) + i\omega_{s})}\right\}\; .
\label{Sig11}
\end{eqnarray}
Since in strong coupling $f(E({\bf k}))\to 0$, $u^2_{{\bf k}}\to 1$,
and $v^2_{{\bf k}} \to 0$, the second and fourth term within braces on the 
right-hand side of the Matsubara expression  (\ref{Sig11}) may be 
dropped. The simplified expression (\ref{Sigma-11-strong-coupling}) then 
results 
from Eq.~(\ref{Sig11}). In the strong-coupling limit, one would then be 
tempted to perform the analytic continuation $i\omega_s\to z$ directly from the
expression (\ref{Sigma-11-strong-coupling}). Care must, however, be exerted on
this point
since {\em the processes of taking the strong-coupling limit and performing 
the analytic continuation may not commute\/}. By performing the analytic 
continuation $i\omega_s \to z$ directly in Eq.~(\ref{Sig11}) one, in fact, 
obtains
two additional terms with respect to the analytic continuation of 
Eq.~(\ref{Sigma-11-strong-coupling}). These two additional terms cannot be 
dropped {\em a priori} by the presence of the small factor 
$v^2_{{\bf q} - {\bf k}}$ in the strong-coupling limit, because for real $z$ 
the corresponding energy denominators may vanish. Retaining properly 
these two additional terms indeed affects in a qualitative way the spectral 
function $A({\bf k},\omega)$ in the strong-coupling limit, as discussed in 
Sec.~III.

With the expression obtained by the analytic continuation $i\omega_s \to z$ 
of Eq.~(\ref{Sig11}), one can prove explicitly that 
$\Sigma_{11}({\bf k},z)$ is analytic off the real axis and vanishes like 
$z^{-1}$ along any straight line parallel to the real axis with 
${\rm Im}\, \, z \neq 0$, and that ${\rm sgn} [{\rm Im}\,\Sigma_{11}({\bf k},z)]=
-{\rm sgn} [{\rm Im}\,\, z]$. 
In this way, the properties of the function $\Sigma_{11}({\bf k},z)$, 
required to obtain the retarded Green's function (\ref{G-R-compact}) on the 
real axis, are explicitly verified in the strong-coupling limit.

Once the retarded Green's function has been obtained in the form 
(\ref{G-R-compact}) according to the above prescriptions, its imaginary part
defines the spectral function 
\begin{equation}
A({\mathbf k},\omega) \equiv - (1/\pi)
\mathrm{Im} \, G^{R}({\mathbf k},\omega)
\label{akw}
\end{equation}
which will be calculated numerically in Sec.~III
for a wide range of temperatures and couplings. In the Appendix, it is shown 
at a formal level that $A({\mathbf k},\omega)$ satisfies the sum rule   
(\ref{sum-rule-G-R}). This sum rule will be considered an important test for 
the
numerical calculations of Sec.~III. To this end, it is necessary to prove that
the sum rule (\ref{sum-rule-G-R}) holds even for our approximate theory based
on the Dyson's equation (\ref{Dyson-equation}).

To prove the sum rule (\ref{sum-rule-G-R}) for the approximate theory, it is 
sufficient that the approximate $G_{11}({\bf k},z)$ (from which the retarded
Green's function (\ref{G-R-compact}) results when $z=\omega + i \eta$) 
behaves like $z^{-1}$ for large $|z|$. This property is verified by our 
theory, as shown above. 
As a consequence:    
\begin{eqnarray}
\int_{-\infty}^{+\infty} \! d\omega \, A({\mathbf k},\omega) =
 - \frac{1}{\pi}\, 
\mathrm{Im} \left[ \int_{-\infty}^{\infty} \! d\omega \, 
G^{R}({\mathbf k},\omega)\right]\nonumber\\
= - \frac{1}{\pi}\, 
\mathrm{Im} \left[ - \oint_{C} \! d\omega \, G_{11}({\mathbf k},z)
\right] = 1 \;            \label{sum-rule-A}
\end{eqnarray}
where the contour $C$ is a half-circle in the upper-half complex plane with
center in the origin, large radius
(such that the approximation $G_{11}({\mathbf k},z) \sim  z^{-1}$ is
valid), and counterclockwise direction.

Finally, the analytic continuation of the anomalous
Matsubara single-particle Green's function (\ref{G-12-Matsubara}) can be 
obtained 
by following the same procedure adopted for the normal Green's function
(\ref{G-11-Matsubara}).
One writes for the retarded anomalous Green's 
function
\begin{eqnarray}
&& F^R({\bf k},\omega)=
\Delta [(\omega + i\eta -\xi({\mathbf k})-
\Sigma_{11}^R({\mathbf k},\omega)) \nonumber\\
&&\times  (\omega + i\eta+\xi({\mathbf k})+
\Sigma_{11}^R(-{\mathbf k},-\omega)^*) - \Delta^{2}]^{-1}\; . 
\label{FRet} 
\end{eqnarray}
in the place of Eqs.~(\ref{G-R-compact}) and~(\ref{sigma-tilde-real}).
In this case, the analytic properties of $\Sigma_{ii}({\mathbf k},z)$ 
($i=1,2$) discussed above imply that $G_{12}({\mathbf k},z) \sim - 
\Delta/ z^{2}$ asymptotically for large $|z|$.
As a consequence, the imaginary part of $F^R({\bf k},\omega)$
\begin{equation}
B({\mathbf k},\omega) \equiv - (1/\pi)
{\mathrm Im}\,F^{R}({\mathbf k},\omega)
\end{equation}
satisfies the two following sum rules:
\begin{equation}
\int_{-\infty}^{+\infty} \! d\omega \, B({\mathbf k},\omega) \, = \, 0
\label{sum-rule-B-1}
\end{equation}
and
\begin{equation}
\int_{-\infty}^{+\infty} \! d\omega \, B({\mathbf k},\omega) \, \omega \, =
\, - \, \Delta \,\, .      \label{sum-rule-B-2}
\end{equation}
These sum rules can be verified by introducing the contour $C$ as in
Eq.~(\ref{sum-rule-A}).
Note again that these sum rules (which are proved on general grounds in the 
Appendix  for the exact anomalous retarded
single-particle Green's function) follow here from our approximate form of
$F^{R}({\mathbf k},\omega)$ only on the basis
of the properties of analyticity.
Verifying numerically the sum rules (\ref{sum-rule-A}),
(\ref{sum-rule-B-1}), and (\ref{sum-rule-B-2}) at
any coupling and temperature will, in practice, constitute an important
check on the validity of the above procedure for the analytic continuation.

An additional numerical check on the validity of the whole procedure at
intermediate-to-weak coupling will be provided by
the merging of the results, obtained by calculating the spectral
function $A({\mathbf k},\omega)$ when approaching
$T_{c}$ from below, with the results previously obtained in the normal
phase\cite{PPSC-02} when approaching $T_{c}$ from above.

\section{Numerical results and discussion}
In this section we present the numerical results based on the formal theory
developed in Sec.~II. Specifically,
in Sec.~IIIA we present the results obtained by solving the 
coupled equations~(\ref{n-G-11}) and (\ref{BCS-gap_equation}) for the 
order parameter and chemical potential. Section IIIB deals instead 
with the numerical calculation of the spectral function (\ref{akw})  
in the broken-symmetry phase, over the whole coupling range from weak to 
strong. 
\subsection{Order parameter and chemical potential}

Before presenting the numerical results for $\Delta$ and $\mu$, it is worth 
outlining briefly the numerical procedure we have adopted.

At given temperature and coupling, the coupled equations~(\ref{n-G-11}) 
and (\ref{BCS-gap_equation}) for the unknowns $\Delta$ and $\mu$ are solved 
via the Newton's method. This requires knowledge of the self-energy 
$\Sigma_{11}({\bf k},\omega_{s})$ of Eq.~(\ref{Sigma-broken-11}), with 
$\Gamma_{11} (q)$ obtained from
Eqs.~(\ref{Gamma-solution})-(\ref{B-definition}).
[As anticipated, in the numerical calculations we neglect $\Sigma_{12}^L$ in 
comparison to $\Sigma_{12}^{{\rm BCS}}$, 
since inclusion of $\Sigma_{12}^L$ is not required to recover the 
Bogoliubov results in the strong-coupling limit, as shown in Sec.~IID.]    

To this end, the frequency sums in 
Eqs.~(\ref{A-definition}) and (\ref{B-definition}) 
are evaluated analytically, while the remaining wave-vector integral
is calculated numerically by the Gauss-Legendre method.
In particular, the radial wave-vector integral extending up to infinity is 
partitioned into an inner and an outer region, with the transformation 
$|{\bf p}|\to 1/|{\bf p}|$  
exploited in the outer region.
     
The bosonic frequency sum in Eq. (\ref{Sigma-broken-11}) requires 
special care, owing to its slow convergence and the lack of an intrinsic 
energy cutoff within our continuum model. 
We have accordingly partitioned this frequency sum into three regions, 
separated by the frequency scales $\Omega_{c_1}$ and $\Omega_{c_2}$ (with 
$0<\Omega_{c_1}<\Omega_{c_2})$. 
For $|\Omega_{\nu}|<\Omega_{c_1}$, the frequency sum is calculated explicitly.
For $\Omega_{c_1}<|\Omega_{\nu}|<\Omega_{c_2}$, the frequency sum is 
approximated  with great accuracy by the corresponding numerical integral, 
owing to the slow dependence of $\Gamma_{11}$ on $\Omega_{\nu}$. 
Finally, the tail of the frequency sum for $\Omega_{c_2}<|\Omega_{\nu}|$ 
(where the asymptotic expression~(\ref{gammasy}) yields $\Gamma_{11} 
\propto (i\Omega_{\nu})^{-1/2}$) is evaluated 
analytically. Typically, $\Omega_{c_1}$ is taken of the order of  
the largest among the energy scales $|\omega_{s}|,\Delta,|\mu|,{\bf q}^2/(2m)$,
and ${\bf k}^2/(2m)$; $\Omega_{c_2}$ is then taken at least ten times  
$\Omega_{c_1}$. 
It turns out that it is most convenient to apply this procedure to the 
frequency sum in Eq.~(\ref{Sigma-broken-11}) after the integration over the two
angular variables of the wave vector ${\bf q}$ has been performed 
analytically; the remaining radial wave-vector integration 
is then performed numerically, with a 
cutoff much larger than the wave-vector scales
$|{\bf k}|, \sqrt{2m \Delta}$, and $\sqrt{|\mu|}$.

\begin{figure}
\includegraphics[scale=0.7]{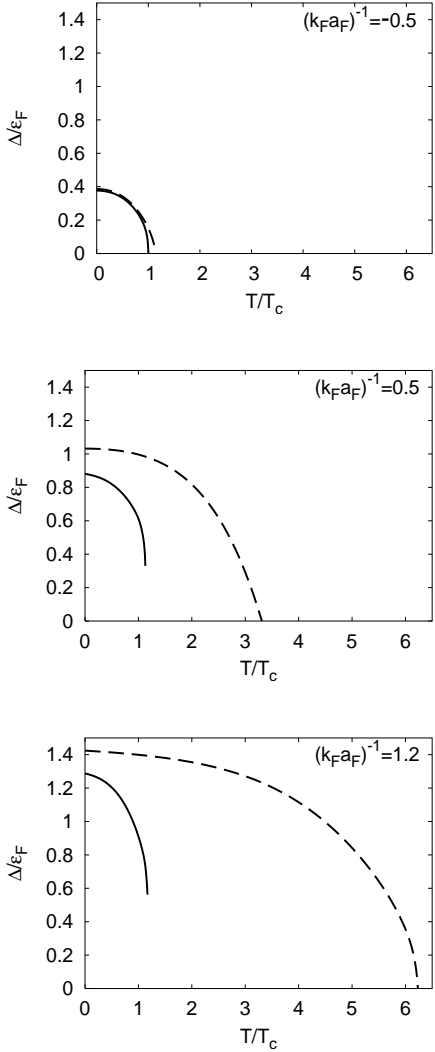}
\vspace{.25truecm}
\caption{Order parameter $\Delta$ (in units of $\epsilon_F$) vs 
temperature (in units of $T_c$) for different values of the coupling
$(k_F a_F)^{-1}$. 
Results obtained by the inclusion of fluctuations (full lines) are compared 
with mean-field results (dashed lines). }
\label{gapvsT}
\end{figure}

Finally, the frequency sum in the particle number equation (\ref{n-G-11}) is 
evaluated by adding and subtracting the BCS Green's function 
${\cal G}_{11}$ on the right-hand side of that equation, in order to 
speed up the numerical convergence.
Matsubara frequencies are here summed numerically up to a cutoff frequency,
beyond 
which the sum is approximated by the corresponding numerical integral. 
The radial part of the wave-vector integral in Eq.~(\ref{n-G-11}) 
is also calculated numerically up to a cutoff scale beyond which a power-law 
decay sets in, so that the contribution from the tail can be calculated 
analytically.

\begin{figure}
\includegraphics[scale=0.5]{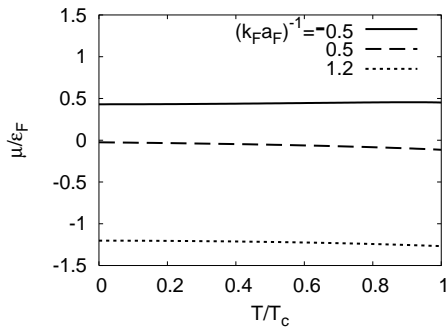}
\vspace{.25truecm}
\caption{Chemical potential $\mu$ (in units of $\epsilon_F$) vs 
temperature (in units of $T_c$), for  
the same values of the coupling $(k_Fa_F)^{-1}$ as in Fig.~\ref{gapvsT}.} 
\label{muvsT}
\end{figure}

\begin{figure}
\includegraphics[scale=0.5]{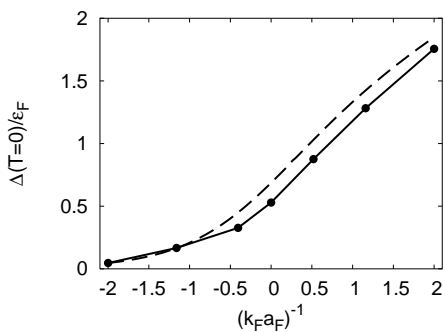}
\vspace{.25truecm}
\caption{Order parameter $\Delta$ at $T=0$ (in units of $\epsilon_F$) vs  
the coupling $(k_Fa_F)^{-1}$ .     
Results obtained by the inclusion of fluctuations (full line) are compared 
with mean-field results (dashed line).}
\label{gapt0}
\end{figure}

\begin{figure}
\includegraphics[scale=0.5]{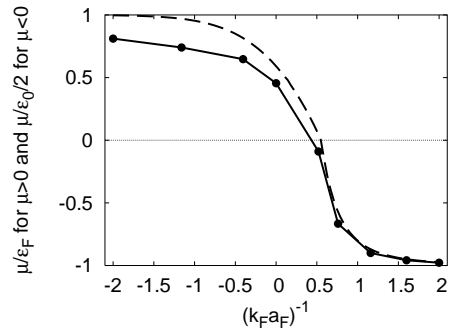}
\vspace{.25truecm}
\caption{Chemical potential $\mu$ at $T=0$ (in units of $\epsilon_F$ for 
$\mu >0$ and 
of $\epsilon_0/2$ for $\mu<0$) vs the coupling $(k_Fa_F)^{-1}$.
Results obtained by the inclusion of fluctuations (full line) are compared 
with mean-field results (dashed line).}
\label{mut0}
\end{figure}

With the above numerical prescriptions, we have obtained the behavior of 
$\Delta$ 
and $\mu$ vs temperature and coupling reported in 
Figs.~\ref{gapvsT}-\ref{mut0}.

Specifically, Fig.~\ref{gapvsT} shows the order parameter
$\Delta$ vs temperature for different couplings 
[$(k_F a_F)^{-1}=-0.5,0.5,1.2$, from top to bottom], in the window 
$-1 \lesssim (k_F a_F)^{-1}\lesssim +1$ where the crossover from weak to
strong coupling is exhausted. Comparison is made with the corresponding
curves obtained within mean field (dashed lines), when the BCS Green's
function ${\mathcal G}_{11}$ enters Eq.~(\ref{n-G-11}) in 
the place of the dressed $G_{11}$. In these plots, the temperature is  
normalized with respect to the critical temperature $T_c$ for the given 
coupling. This comparison shows that fluctuation corrections on top of 
mean field get progressively important at given coupling as the 
temperature is raised toward $T_c$. Close to $T_c$, fluctuation 
corrections become even more important upon approaching the strong-coupling 
limit.
Near zero temperature, on the other hand, fluctuation corrections become 
negligible when approaching strong coupling. This confirms the expectation 
that, near zero temperature, the BCS mean field should be rather 
accurate both in the weak- and strong-coupling limits~\cite{Leggett-80}. 

Note from Fig.~\ref{gapvsT} that $\Delta$ jumps discontinuously close to the 
critical temperature when
fluctuations are included on top of mean field. This jump becomes more evident
as the coupling is increased. It reflects an 
analogous behavior of the condensate density near the critical temperature as
obtained by the Bogoliubov theory for point-like bosons~\cite{Luban} . In the 
present theory this jump is carried over to the composite bosons, even 
at fermionic couplings [as in the middle panel of Fig.~\ref{gapvsT} ] when 
the composite bosons
are not yet fully developed. When the fermionic coupling increases beyond 
the values reported in Fig.~\ref{gapvsT}, however, the residual interaction 
between the composite bosons decreases further and the jump becomes 
progressively smaller.  
More refined theories for point-like bosons (see, 
e.g.,~Ref.~\onlinecite{griffin}) remove the jump of the bosonic condensate 
density, which thus should be considered as an artifact of the 
Bogoliubov approximation.
Apart from this jump, note that when the temperature is decreased 
below $T_c$ the order parameter $\Delta$ grows more rapidly with the 
inclusion of fluctuations than within mean field. 

Figure~\ref{muvsT} shows the chemical potential $\mu$ vs temperature for 
the same coupling values of Fig.~\ref{gapvsT}. Note that in weak coupling the 
chemical potential decreases slightly upon moving deep in the superconducting 
phase from 
$T_c$ to $T=0$, in agreement with the BCS behavior. In strong coupling 
the opposite occurs, reflecting the behavior
of the bosonic chemical potential $\mu_B=2\mu+\epsilon_0$ within the 
Bogoliubov theory. It should be, however, mentioned that with improved 
bosonic approximations~\cite{griffin}, the bosonic chemical potential would 
rather decrease upon entering the condensed phase.

Figure~\ref{gapt0} shows the order parameter $\Delta$ at zero temperature 
(full line) and the corresponding mean-field value (dashed line)
vs the coupling $(k_F a_F)^{-1}$.
While $\Delta$ increases monotonically in absolute value 
from weak to strong coupling (as expected on physical grounds), the relative 
importance of the fluctuation corrections to the order parameter at zero 
temperature (over and above mean field) 
reaches a maximum in the intermediate-coupling region, 
never exceeding about 30\%. 
This results confirms again that the BCS mean field is a reasonable 
approximation to 
the ground state for all couplings.

Figure~\ref{mut0} shows the chemical potential $\mu$ at zero 
temperature vs the coupling parameter $(k_F a_F)^{-1}$. The results obtained 
by the inclusion of fluctuations (full lines) are compared with  mean field 
(dashed lines). Even for this thermodynamic quantity the fluctuation 
corrections to the mean-field results appear to be not too important at zero 
temperature.

Note, finally, that the values for $\Delta$ and $\mu$ obtained from our theory
at $T=0$ with the coupling value $(k_F a_F)^{-1}=0$ are in remarkable 
agreement with a recent Quantum Monte Carlo calculation~\cite{carlson} 
performed for the same coupling. Our calculation yields, in fact, 
$\Delta/\epsilon_F=0.53$ and $\mu/\epsilon_F = 0.445$, to be compared with the
values $\Delta/\epsilon_F=0.54$ and $\mu/\epsilon_F = 0.44 \pm 0.01$ of
Ref.~\onlinecite{carlson}. [In contrast, BCS mean field yields 
$\Delta/\epsilon_F=0.69$ and $\mu/\epsilon_F = 0.59$.]

In summary, the above results have shown that, for thermodynamic quantities 
like $\Delta$ and $\mu$, fluctuation corrections to mean-field values 
in the broken-symmetry phase are important only as far as the temperature
dependence is concerned, while at zero temperature the mean-field results are 
reliable.

\subsection{Spectral function}

For a generic value of the coupling, calculation of the imaginary part of the 
retarded self-energy 
${\rm Im} \Sigma_{11}^R({\bf k},\omega)=-h({\bf k},\omega)$ (with 
$h({\bf k},\omega)$ given by Eq.~$(\ref{imsig11})$) requires us to obtain the 
imaginary part of the particle-particle ladder $\Gamma_{11}^R$ on the 
real-frequency axis, as determined by  
the formal replacement  $i\Omega_\nu \rightarrow \omega+i\eta$ in the 
Matsubara expressions (\ref{Gamma-solution})-(\ref{B-definition}).
After performing the frequency sum therein, the wave-vector integrals of
Eqs.~(\ref{A-definition}) and (\ref{B-definition}) for the functions 
$\chi_{ij}({\bf q},i\Omega_\nu \rightarrow \omega+i\eta)$ ($(i,j)=1,2$) are
 evaluated  numerically, by exploiting the properties of the delta function 
for the imaginary part and keeping a finite albeit small value of 
$\eta \, (=10^{-8}\sqrt{\mu^2+\Delta^2})$ for the real part.

\begin{figure}
\includegraphics[scale=0.7]{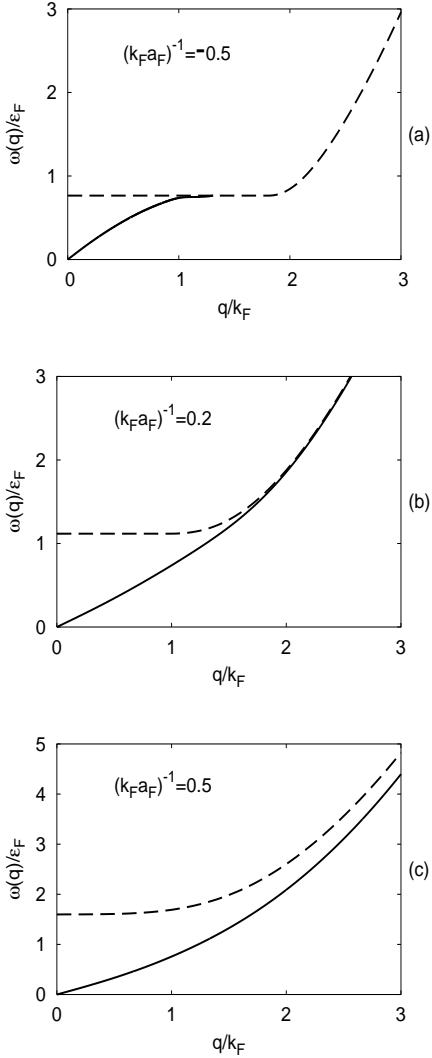}
\vspace{.25truecm}
\caption{Dispersion $\omega({\bf q})$ of the pole of 
$\Gamma^R_{11}({\bf q},\omega)$ at 
$T=0$ (full lines) and boundary of the particle-particle 
continuum (dashed lines) for three 
characteristic couplings.}
\label{gamma}
\end{figure}

Direct numerical calculation of the imaginary part of the particle-particle 
ladder
fails, however, when this part has the structure of a delta function for real
$\omega$ at given ${\bf q}$. This occurs when the determinant in the 
denominator of Eq.~(\ref{Gamma-solution}) vanishes for real $\omega$.
To deal with this delta function,
let's first consider the case $T=0$ for which three cases can be 
distinguished, according to: (i) $a_F<0$ and $\mu>0$ 
(weak-to-intermediate 
coupling); (ii) $a_F>0$ and $\mu>0$ (intermediate coupling); (iii) $a_F>0$
and $\mu<0$ (intermediate-to-strong coupling). The curves $\omega({\bf q})$ 
where the (analytic continuation of the) determinant in the denominator of 
Eq.~(\ref{Gamma-solution}) vanishes are shown  (full lines) for these three 
cases in Figs.~\ref{gamma} (a), (b), and (c), respectively. In these figures 
we also show the boundaries (dashed lines) delimiting the 
particle-particle continuum, where the imaginary 
part of the particle-particle ladder is nonvanishing and regular
(in the sense that it does not have the structure of a delta function). At 
finite temperature, the sharp boundary of the 
particle-particle
continuum smears out, owing to the presence of Fermi functions 
after performing the sum over the Matsubara frequencies in 
Eqs.~(\ref{A-definition}) and~(\ref{B-definition}). The Fermi functions 
produce, in fact, a finite 
(albeit exponentially small with temperature) imaginary part of the 
particle-particle ladder also below the (dashed) boundaries of 
Fig.~\ref{gamma},
resulting in a Landau-type damping of the Bogoliubov-Anderson mode 
$\omega({\bf q})$. 
In addition, the finite imaginary part broadens the delta-function structure
centered at the curves $\omega({\bf q})$ of Fig.~\ref{gamma}, turning it into
 a Lorentzian function. In practice, our numerical calculation takes advantage 
of this broadening occurring at finite temperature, and deals with smooth 
Lorentzian functions instead of the delta-function peaks.~\cite{footnotegamma} 

As a further consistency check on our numerical calculations, we have 
sistematically verified that
the three sum rules (\ref{sum-rule-A}), (\ref{sum-rule-B-1}), 
(\ref{sum-rule-B-2}) are satisfied
within numerical accuracy, for all temperatures and couplings we have 
considered.

\begin{figure}
\includegraphics[scale=0.7]{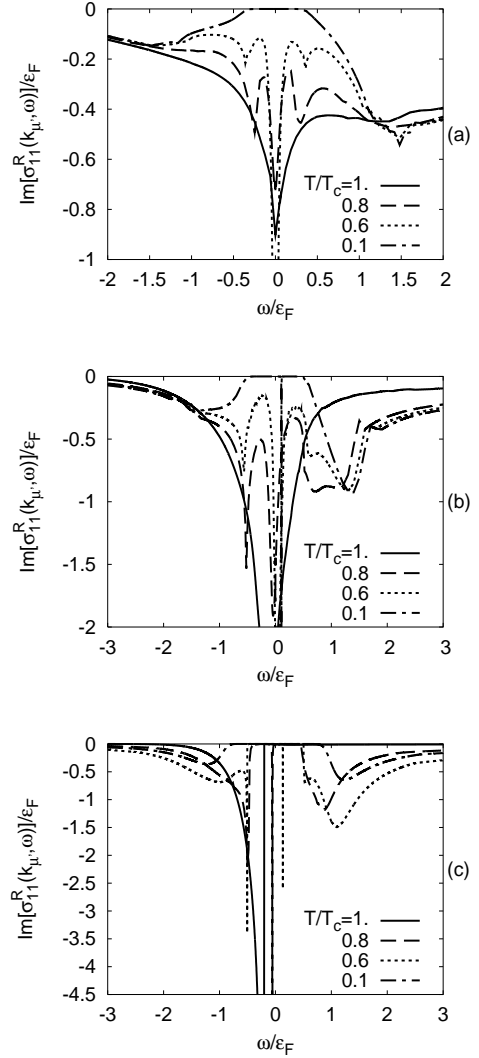}
\vspace{.25truecm}
\caption{Imaginary part of the self-energy $\sigma^R_{11}$ for 
$|{\bf k}|=k_{\mu'}$ vs frequency (in units of $\epsilon_F$) at 
different temperatures for the coupling values 
$(k_F a_F)^{-1}=$ -0.5 (a), 0.1 (b), and 0.5 (c).}
\label{sigmaim}
\end{figure}

\begin{figure}
\includegraphics[scale=0.7]{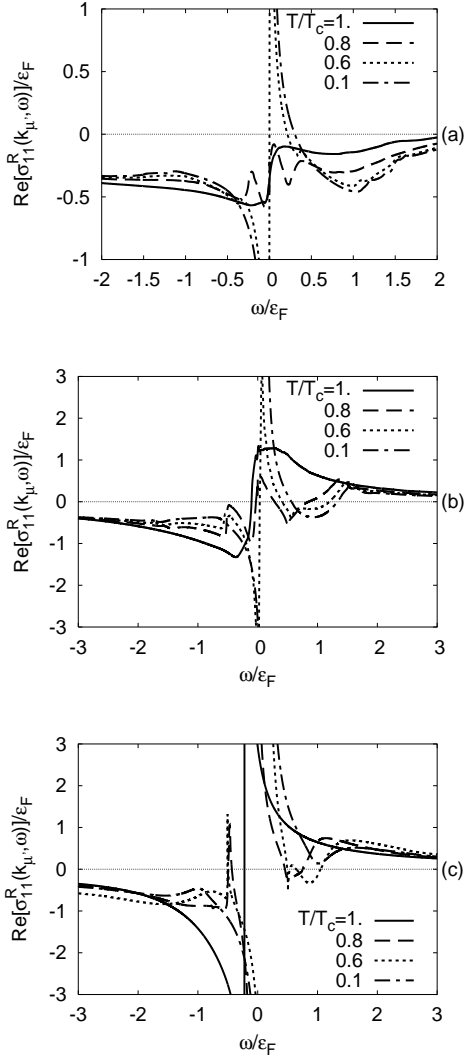}
\vspace{.25truecm}
\caption{Real part of the self-energy $\sigma^R_{11}$ for 
$|{\bf k}|=k_{\mu'}$ vs frequency
(in units of $\epsilon_F$) at different temperatures for the coupling 
values $(k_F a_F)^{-1}=$ -0.5 (a), 0.1 (b), and 0.5 (c).}
\label{sigmareal}
\end{figure}

The imaginary and real parts of the retarded self-energy 
$\sigma_{11}^R({\bf k},\omega)$ obtained from 
 Eq.~(\ref{sigma-tilde-real})  
are shown, respectively, in Figs.~\ref{sigmaim} and~\ref{sigmareal} as 
functions of frequency at different temperatures and for different couplings 
(about the crossover region of interest). The magnitude of the wave vector 
${\bf k}$ is taken in Figs.~\ref{sigmaim} and~\ref{sigmareal} at a special 
value (denoted by $k_{\mu'}$), which is identified from the behavior of
the ensuing spectral function $A({\bf k},\omega)$ when performing a scanning
over the wave vector (see Fig.~12 below).  Accordingly, $k_{\mu'}$ is chosen 
to minimize the gap in the spectral function, in agreement with a standard 
procedure in the ARPES literature.
On the weak-coupling side (when the the self-energy shift $\Sigma_0$ discussed
in Sec.~IIB is included in our calculation), $k_{\mu'}$ coincides with 
$\sqrt{ 2 m (\mu - \Sigma_0)}$. On the strong-coupling side (when 
$\mu$ becomes negative) one takes instead $k_{\mu'} = 0$.  

\begin{figure}
\includegraphics[scale=0.7]{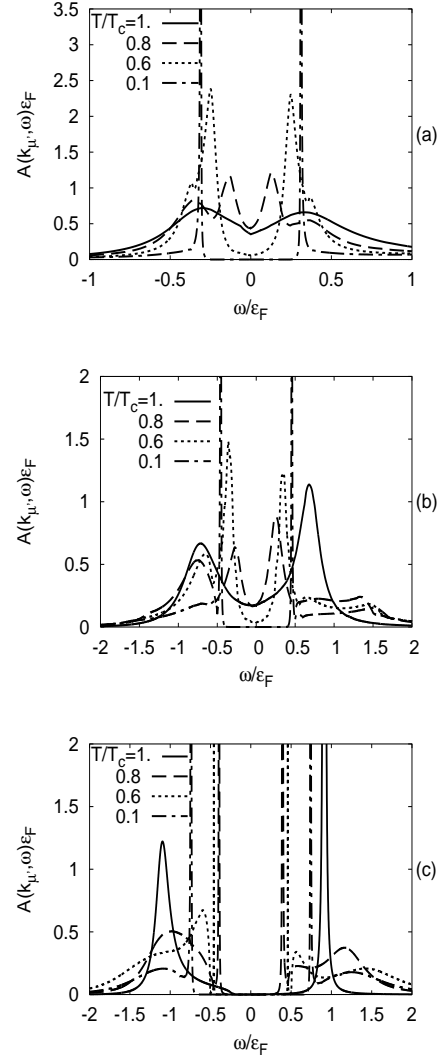}
\vspace{.25truecm}
\caption{Spectral function for $|{\bf k}|=k_{\mu'}$ vs frequency
(in units of $\epsilon_F$) at different temperatures for the coupling values
$(k_F a_F)^{-1}=$ -0.5 (a), 0.1 (b), and 0.5 (c).}
\label{spectT}
\end{figure}

For all couplings here considered, the progressive evolution 
found in $A({\bf k},\omega)$ (from the 
presence of a pseudogap about $\omega=0$ at $T_c$ to the occurrence of a 
superconducting gap near zero temperature) stems from the interplay  
of the two contributions in Eq.~(\ref{sigma-tilde-real}) to the imaginary part
of $\sigma^R_{11}({\bf k},\omega)$ about $\omega=0$. Specifically, for  
intermediate-to-weak coupling (with $\mu>0$) the first term on the 
right-hand side of Eq.~(\ref{sigma-tilde-real}) (which is responsible for the 
pseudogap suppression in $A({\bf k},\omega)$ at $T_c$) would produce a narrow
peak structure in $A(k_{\mu'},\omega)$ about $\omega=0$ upon lowering $T$, 
since ${\rm Re} \Sigma_{11}^{R}(k_{\mu'},\omega)-\Sigma_0$ 
vanishes while $|{\rm Im}\,\Sigma_{11}^R(k_{\mu'},\omega)|$ becomes 
progressively smaller. The presence of the second 
term on the right-hand-side of Eq.~(\ref{sigma-tilde-real}), however, gives 
rise to a narrow peak in ${\rm Im}\, \sigma^R_{11}(k_{\mu'},\omega)$ about 
$\omega=0$, as seen from Fig.~\ref{sigmaim} (a), resulting in a depression 
of $A(k_{\mu'},\omega)$ about $\omega=0$.   
[This occurs barring a small temperature range close to $T_c$, where the 
second term on the right-hand-side of Eq.~(\ref{sigma-tilde-real}) is not yet 
well developed.]
At larger couplings (when $\mu < 0 $), the first term on the right-hand-side
of Eq.~(\ref{sigma-tilde-real}) would not produce a peak in 
$A({\bf k}=0,\omega)$ about $\omega=0$ upon lowering the temperature, because 
$|\mu|+{\rm Re} \Sigma_{11}^R(k=0,\omega)$ does not correspondingly vanish in 
this case even though ${\rm Im}\,\Sigma_{11}^R({\bf k}=0,\omega)$ does. 
In addition, in this case the second term on the right-hand-side of 
Eq.~(\ref{sigma-tilde-real}) does not produce a peak in $A({\bf k}=0,\omega)$
about $\omega=0$.

\begin{figure}
\includegraphics[scale=0.5]{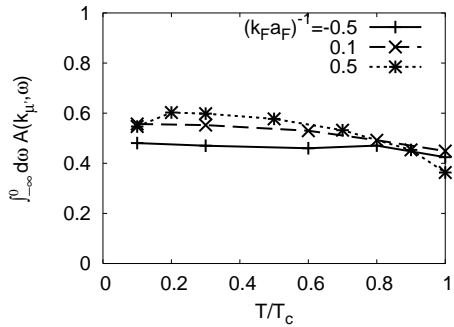}
\vspace{.25truecm}
\caption{Temperature dependence of the total weight of the spectral 
function at negative frequencies for different coupling values about the 
crossover region.}
\label{weightneg}
\end{figure}

Figure~\ref{spectT} shows the resulting spectral function 
$A({\bf k},\omega)$ vs $\omega$ for $|{\bf k}|=k_{\mu'}$  
at different temperatures and couplings. In all cases, at $T_c$
there occurs only {\em a broad pseudogap feature} both for $\omega >0$ and
$\omega <0$. [For photoemission experiments only the case $\omega <0$ 
is relevant, so that we shall mostly comment on this case in the following.] 
A {\em coherent peak} is seen to grow on top of this broad pseudogap feature 
as the temperature is lowered below $T_c$. When zero temperature is 
eventually reached, the pseudogap feature is partially suppressed in favor
of the coherent peak, which thus absorbs a substantial portion of the 
spectral intensity. This 
interplay between the broad pseudogap feature and the sharp coherent peak 
results in a characteristic {\em peak-dip-hump structure\/}, which is best
recognized from the features for weak-to-intermediate coupling.
Generally speaking, this coherent peak (and its corresponding counterpart at
positive frequencies) for intermediate-to-weak coupling is associated 
with the two dips in ${\rm Im}\, \sigma^R_{11}$ symmetrically located 
about zero frequency [cf.~Figs.~\ref{sigmaim} (a) and (b)]. 
In strong coupling, instead, the coherent peak results from a delicate balance
between the real and imaginary parts of $\sigma^R_{11}$ 
near the boundary of the region where ${\rm Im}\,\Sigma^R_{11}=0$.

An interesting fact is that the weights of the negative and positive frequency
parts of the spectrum turn out to be separately (albeit approximatively) 
constant as functions of temperature for given coupling, as shown in 
Fig.~\ref{weightneg} for three characteristic couplings. This implies that, 
for a given coupling, the coherent peak for $\omega < 0$ grows at the 
expenses of the accompanying broad pseudogap feature upon 
decreasing the temperature. 

The result that the total area for {\em negative} $\omega$ should be 
(approximately)
constant as a function of temperature can be realized also from the analytic 
results in the extreme strong-coupling limit discussed in Sec.~IID. 
Taking the analytic continuation of the Matsubara Green's function 
(\ref{G-11-strong-coupling-figo}) (which is 
appropriate in the strong-coupling limit as far as this total area is 
concerned, as it will be shown below)
results, in fact,  in the total weight $\tilde{v}^2({\bf k})$ of the 
$\omega < 0$ region being independent of temperature, since the combination 
$\Delta^2+\Delta_0^2$ entering the expression of $\tilde{E}({\bf k})$ is 
proportional to the total density in this limit 
[cf.~Eq.~(\ref{final-n})].

Returning to Fig.~\ref{spectT}, it is also interesting to comment on the 
positions of the pseudogap feature and the coherent peak as functions of 
temperature for given coupling. The position of the coherent
peak depends markedly on temperature, shifting progressively toward more
negative frequencies as the temperature is lowered. In particular, for 
weak-to-intermediate coupling the position of the coherent peak about coincides
 with (minus) the value of the order parameter $\Delta$. In the 
strong-coupling region (where $\mu < 0$), on the other hand, its position 
is about at $-\sqrt{\Delta^2+\mu^2}$. This remark entails the 
possibility of extracting two important quantities from the temperature 
evolution of the coherent peak in the spectral function: (i) The frequency 
position of this peak when approaching $T_c$ determines whether $\mu$ is 
positive (when the peak position approaches $\omega=0$) or negative (when the 
peak position approaches $-|\mu|$), corresponding to weak-to-intermediate 
coupling and strong coupling, respectively; (ii) In both cases, the 
temperature dependence of the order parameter can be extracted from the 
frequency position of the coherent peak.

The above results for the coherent peak contrast somewhat with the position of
the pseudogap feature by decreasing temperature below $T_c$, also determined 
from Fig.~\ref{spectT}. The broad pseudogap feature does not depend 
sensitively on temperature for all couplings shown in this figure.
This indicates that the 
broad pseudogap feature does not relate to the order parameter below $T_c$.

As far as the spectral function is concerned, one of the key results of our 
theory is thus the presence of {\em two structures\/} (coherent peak and 
pseudogap), which behave rather independently from each other as functions of
temperature and coupling. This result, which is also evidenced by the behavior
of the experimental spectra in tunneling experiments on 
cuprates~\cite{kugler}, originates in our theory from the presence of 
two distinct contributions to the self-energy, namely, the BCS and fluctuation 
contributions of Eq.~(\ref{Dyson-equation}). While the broad 
pseudogap feature at $T<T_c$ develops with continuity from the only feature 
present at $T>T_c$, the coherent peak {\em per se} would be present in a BCS 
approach even in the absence of the fluctuation contribution. This remark, of 
course, does not imply 
that the two contributions to the self-energy of Eq.~(\ref{Dyson-equation})
are totally independent from each other. They both 
depend, in fact, on the value of the order parameter $\Delta$ which is, 
in turn, determined by both self-energy contributions via the chemical 
potential.

\begin{figure}[h]
\includegraphics[scale=0.5]{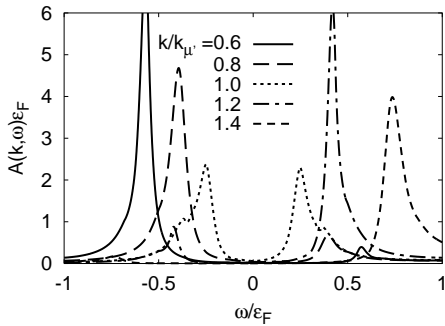}
\vspace{.25truecm}
\caption{Spectral function at different wave vectors about $k_{\mu'}$ for
$T=0.6 T_c$ vs frequency (in units of $\epsilon_F$) for the coupling value 
$(k_F a_F)^{-1}=-0.5$.}
\label{momentumevol}
\end{figure}

A further important feature that can be extracted from our calculation
of the spectral function is the evolution of the coherent peak for 
varying wave vector at fixed temperature and coupling. 
Figure~\ref{momentumevol} reports $A({\bf k},\omega)$ vs $\omega$ for 
different values of the ratio $k/k_{\mu'}$ about unity when 
$(k_F a_F)^{-1}=-0.5$ and $T/T_c=0.6$. Here, $k/k_{\mu'}=1$ identifies the 
underlying Fermi surface that represents the ``locus of minimum gap''. 
When $k/k_{\mu'} < 1$, there is a strong asymmetry between the two coherent 
peaks at $\omega < 0$ and $\omega > 0$, with the peak at $\omega < 0$ 
absorbing most of the total weight. The situation is 
reversed when $k/k_{\mu'} > 1$. When $k/k_{\mu'} = 1$ the spectrum is (about)
symmetric between $\omega$ and $-\omega$. In addition, when following the 
position of the coherent peak at $\omega < 0$ starting from $k/k_{\mu'} < 1$,
one sees that this position moves toward increasing $\omega$, reaches a 
minimum distance from $\omega=0$, and bounces eventually back to more 
negative values of $\omega$.
The value of the minimum distance from $\omega=0$ identifies an energy scale 
$\Delta_m$.
 At the same time,
the weight of the coherent peak at $\omega < 0$ progressively 
decreases for increasing $k/k_{\mu'}$ starting from $k/k_{\mu'} < 1$.
When $k/k_{\mu'}$ becomes larger than unity, the weight of the coherent peak
is transferred from negative to positive frequencies. This situation is 
characteristic of the BCS theory, where only the coherent 
peaks are present without the accompanying broad pseudogap features. 
Our calculation shows that this situation persists also for 
couplings values
inside the crossover region, where the presence of the pseudogap feature is
well manifest due to strong superconducting fluctuations. 
[Sufficiently far from the underlying Fermi surface, the coherent peak and 
the pseudogap feature merge into a single structure, as it is evident from
Fig.~\ref{momentumevol}. In this case, the above as well as the 
following considerations apply to the structure as a whole and not to its 
individual components.]

\begin{figure}[ht]
\includegraphics[scale=0.7]{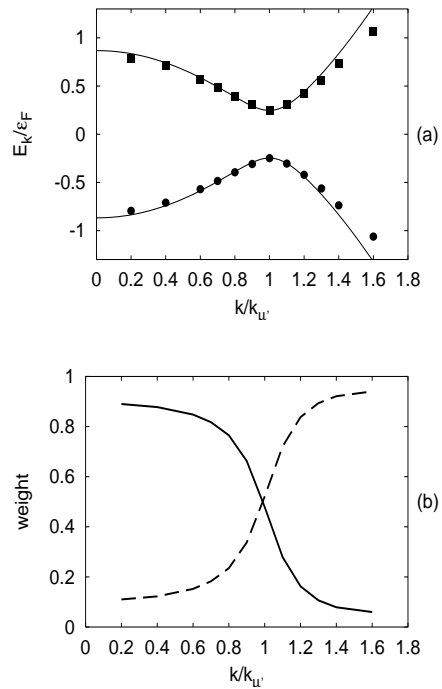}
\vspace{.25truecm}
\caption{(a) Positions of the coherent peaks (in units of $\epsilon_F$) vs 
the wave vector as extracted from Fig.~\ref{momentumevol} . Positive 
(squares) and negative (circles) branches are compared with BCS-like 
dispersions (full lines), as explained in the text. (b) Corresponding weights 
vs the wave vector, with particle-like (full line) and 
hole-like (dashed line) contributions.}
\label{bcslike}
\end{figure}

Figure~\ref{bcslike}(a) summarizes this finding for the dispersion of the
coherent peaks, by showing the positions of the two coherent peaks 
as extracted from Fig.~\ref{momentumevol}  vs $k/k_{\mu'}$. These positions 
are compared with the two branches 
$\pm \sqrt{\xi({\bf k})^2+\Delta_m^2}$ of a BCS-like dispersion, where 
$\Delta_m$ is also identified from Fig.~\ref{momentumevol}. [The value of 
$\Delta_m$ turns out to about coincide with the value of the order parameter 
$\Delta$ at the same temperature, see below.]
The corresponding evolution of the weights of these peaks is shown in 
Fig.~\ref{bcslike}(b), where the characteristic feature of an avoided 
crossing is evidenced. The dispersion of the
positions and weights of the coherent peaks shown in Fig.~\ref{bcslike} compare
favorably with those recently obtained experimentally~\cite{matsui} for 
slightly overdoped Bi2223 samples below the critical temperature 
(for $T/T_c\simeq 0.6$). 

\begin{figure}[ht]
\includegraphics[scale=0.5]{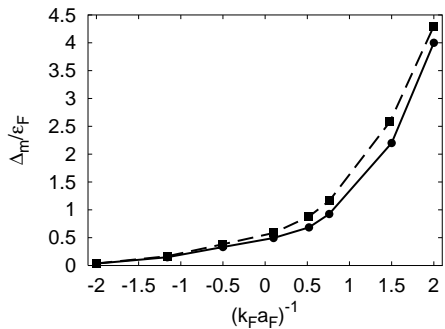}
\vspace{.25truecm}
\caption{Position $\Delta_m$ (in units of $\epsilon_F$) of the quasi-particle 
peak at
$T=0.1 T_c$ vs the coupling $(k_Fa_F)^{-1}$ (full line). The dashed
line corresponds to the value of the order parameter $\Delta$ 
when $\mu > 0$  and of 
$\sqrt{\Delta^2 +{\mu}^2}$ when $\mu < 0$.}
\label{deltam}
\end{figure}

An additional outcome of our calculation is reported in Fig.~\ref{deltam},
where the distance $\Delta_m$ of the coherent peak in $A({\bf k,\omega})$
from $\omega=0$ at $|{\bf k}|=k_{\mu'}$ is compared at low temperature with
the order parameter $\Delta$ when $\mu > 0$ and with $\sqrt{\mu^2+\Delta^2}$ 
when $\mu < 0$. This plot thus compares dynamical and thermodynamic 
quantities. The good agreement between the
two curves confirms our identification of the coherent-peak position in 
$A({\bf k},\omega)$ with the minimum value of the excitations in the 
single-particle spectra according to a BCS-like expression (where 
the value of the order parameter $\Delta$ is, however, obtained by including 
also fluctuation contributions). 

Finally, it is interesting to comment on the strong-coupling 
result~(\ref{G-11-strong-coupling})  
for the diagonal Green's function, with a characteristic double-fraction 
structure. 
The corresponding spectral function $A({\bf k},\omega)$, obtained from that 
expression after
performing the analytic continuation $i\omega_n\to\omega + i\eta$, shows
only a {\em single} feature for $\omega < 0$, with a temperature-independent
position. This contrasts the numerical results we have presented 
[cf. in particular Fig.~\ref{spectT}]. This difference is due to the fact 
that, in our numerical calculation, the analytic continuation has been 
properly 
performed {\em before} taking the strong-coupling limit, as emphasized in 
Sec.~IIE. With this procedure, in fact, the pseudogap structure and the 
coherent peak remain distinct from each other even in the strong-coupling 
limit, without getting lumped into a single feature. 
Such a noncommutativity of the processes of taking the analytic continuation 
and the strong-coupling limit was noted already in a previous 
paper~\cite{PPSC-02} when studying the spectral function above $T_c$. 
More generally, the occurrence of this noncommutativity is expected whenever 
one considers approximate expressions in the Matsubara representation and 
takes the analytic continuation of these expressions to real frequency. 

\begin{figure}[h]
\includegraphics[scale=0.5]{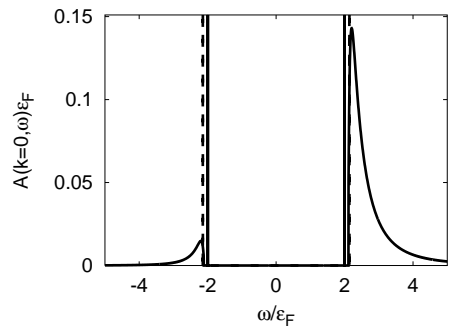}
\vspace{.25truecm}
\caption{Spectral function vs frequency for $(k_F a_F)^{-1}=2.0$ and 
$T/T_c=0.1$, 
obtained by taking alternatively the analytic continuation of $\Sigma_{11}$ 
from the expression (\ref{Sig11}) (full line) or from the expression  
(\ref{Sigma-11-n-prime}) (broken line).}
\label{fignoncomm}
\end{figure}

To make evident the noncommutativity of the two processes, we show in 
Fig.~\ref{fignoncomm} the spectral function $A({\bf k}=0,\omega)$ for
$(k_F a_F)^{-1}=2.0$ and $T/T_c=0.1$, obtained by two alternative methods, 
namely: (i) Using the analytic continuation of the expression (\ref{Sig11}) 
for $\Sigma_{11}$ where $i \omega_s \to \omega + i\eta$ (full line); 
(ii) Taking the strong-coupling expression (\ref{Sigma-11-n-prime}) for 
$\Sigma_{11}$, in which the analytic 
continuation $i\omega_s\to \omega + i \eta$ is performed (broken line). 
Method (i) results in the presence of {\em two} distinct structures in 
$A({\bf k},\omega)$ for $\omega < 0$, corresponding to the coherent 
(delta-like)
peak and the broad pseudogap feature. Method (ii) gives instead a 
{\em single\/} delta-like peak. It is interesting to note that the total 
spectral weight of the two peaks for $\omega < 0$ obtained by method (i) 
(=0.049 for the coupling of Fig.~\ref{fignoncomm}) about coincides with the 
weight of the delta-like peak (=0.044) obtained by method (ii). [We have 
verified that this correspondence between the spectral weights persists also 
at stronger couplings.]

These remarks 
explain the occurrence of a single feature in the spectral function as 
obtained by a different theory based on a preformed-pair 
scenario~\cite{levin98}. In that theory, a single-particle Green's function 
with a double-fraction structure is considered in the 
Matsubara representation for any coupling, and correspondingly a single 
feature in the 
spectral function is obtained for real frequencies~\cite{footnotelevin}. 
Our theory shows instead the appearance of two distinct energy 
scales (pseudogap and order parameter) in the spectral function below $T_c$.

We are thus led to conclude that the occurrence of two distinct energy scales 
below $T_c$ in photoemission and tunneling spectra should not be 
necessarily associated with the presence of an ``extrinsic'' pseudogap due to 
additional non-pairing mechanisms, as sometimes reported in the 
literature~\cite{levin02}.   
\section{Concluding remarks}

In this paper, we have extended the study of the BCS-BEC crossover to finite
temperatures below $T_c$. This has required us to include (pairing) 
fluctuation effects 
in the broken-symmetry phase on top of mean field. Our approximations have 
been conceived to describe both a system of superconducting fermions in weak 
coupling and a system of condensed composite bosons in strong coupling, via
the simplest theoretical approaches valid in the two limits. These are the BCS 
mean field (plus superconducting fluctuations) in weak coupling and the 
Bogoliubov approximation in strong coupling. To this end, analytic results 
have been specifically obtained in strong coupling from our general 
expression of the fermionic self-energy. 

Results of numerical calculations have been presented both for thermodynamic
and dynamical quantities. The latter have
been defined by a careful analytic continuation in the frequency domain.
In this context, a noncommutativity of the analytic continuation and 
the strong-coupling limit has been pointed out.

Results for thermodynamic quantities (like the order parameter and chemical 
potential) have shown that the effects of pairing fluctuations over and above 
the BCS
mean field become essentially irrelevant in the zero-temperature limit, 
even in strong coupling. Results for a dynamical quantity like 
$A({\bf k},\omega)$ have shown, in addition, that two structures 
(a broad pseudogap feature that survives above $T_c$ and a strong coherent
peak which emerges only below $T_c$) are present simultaneously, and that their
temperature and coupling behaviors are rather (even though not completely) 
independent from each other. 

These features produced in the spectral function by our theory originate from
 a totally {\em intrinsic\/} effect, namely, the occurrence of a strong 
attractive interaction (irrespective of its origin). Additional features 
produced by other {\em extrinsic\/} effects could obviously be added on top of 
the intrinsic effects here considered. 

Similar results have recently been obtained in Ref.~\onlinecite{domanski}, 
using 
a boson-fermion model for precursor pairing below $T_c$. In that reference, 
a two-peak structure for $A({\bf k},\omega)$ has also been obtained, although 
with 
a self-energy correction introduced by a totally different method. 

The attractive interaction adopted in this paper is 
the simplest one that can be considered, depending on a single 
parameter only. Detailed comparison of the results of this theory with 
experiments on cuprates would then require one to specify the dependence of 
this effective parameter on temperature and doping. 

The simplified model that we have adopted in this paper should instead be 
considered realistic enough for studying theoretically the BCS-BEC crossover 
for Fermi atoms in a trap. The occurrence of this crossover in these systems 
is being  rather actively studied experimentally at 
present.~\cite{expcross} In this case,
the calculation should also take into account the external trapping 
potential by considering, e.g., a local version of our theory with local 
values of the density and chemical potential in the trap.~\cite{trap}.

\acknowledgments

Financial support from the Italian MIUR under contract COFIN 2001
Prot.2001023848 is gratefully acknowledged.
\appendix
\section{Analytic continuation for the fermionic retarded
single-particle Green's functions
and sum rules below the critical temperature}

In this appendix, we extend {\em below\/} the critical temperature a standard 
procedure for obtaining {\em at a formal level\/} the fermionic retarded 
single-particle 
Green's functions via analytic continuation from their Matsubara counterparts.
This is done in terms of the Lehmann representation \cite{FW} and
of the Baym-Mermin theorem\cite{Baym-Mermin-61}.
In this context, besides the usual sum rule that holds also above the 
critical temperature \cite{FW}, we will
obtain two additional sum rules that hold specifically below the critical
temperature.

The results proved in this appendix hold \emph{exactly\/}, irrespective of the
approximations adopted for the Matsubara self-energy.
To satisfy the above three sum rules with an approximate choice of the
self-energy, however, it is \emph{not\/} required for the ensuing 
approximation to the fermionic single-particle Green's functions to
be ``conserving'' in the Baym's sense \cite{Baym-62}.
Rather, it is sufficient that the analytic continuation from the Matsubara
frequencies to the real frequency axis is taken properly, as demonstrated 
in Sec.~IIE with the specific choice (\ref{total-self-energy})
of the self-energy.

We begin by considering the fermionic ``normal'' and ``anomalous''
\emph{retarded\/} single-particle Green's
functions in the broken-symmetry phase, defined respectively by
\begin{eqnarray}
G^{R}({\mathbf r},t;{\mathbf r'},t') =  - i \theta(t-t') 
      \langle \left\{
\psi_{\uparrow}({\mathbf r},t),\psi_{\uparrow}^{\dagger}({\mathbf r'},t')
\right\} \rangle   \label{G-retarded}  \\
F^{R}({\mathbf r},t;{\mathbf r'},t')  =  - i  \theta(t-t') 
      \langle \left\{
\psi_{\uparrow}({\mathbf r},t),\psi_{\downarrow}({\mathbf r'},t') \right\} 
\rangle\, .   \label{F-retarded}
\end{eqnarray}
Here, $\theta(t)$ is the unit step function, $\psi_{\sigma}({\mathbf r},t)$
is the fermionic field operator
with spin $\sigma=(\uparrow,\downarrow)$ at position ${\mathbf r}$ and
(real) time $t$ (such that
$\psi_{\sigma}({\mathbf r},t) = \exp (iKt) \psi_{\sigma}({\mathbf r}) \exp
(-iKt)$ with $ K = H - \mu N$
in terms of the system Hamiltonian $H$ and the particle number $N$), the braces
represent an anticommutator, and
$\langle\cdots\rangle$ stands for the grand-canonical thermal average.

The Matsubara counterparts of (\ref{G-retarded}) and (\ref{F-retarded}) are
similarly defined by
\begin{eqnarray}
G({\mathbf r},\tau;{\mathbf r'},\tau') & = & - 
\langle T_{\tau} \left[ \psi_{\uparrow}({\mathbf r},\tau) 
\psi_{\uparrow}^{\dagger}({\mathbf r'},\tau') \right] \rangle
\label{G-Matsubara}  \\
F({\mathbf r},\tau;{\mathbf r'},\tau') & = & - 
\langle T_{\tau} \left[ \psi_{\uparrow}({\mathbf r},\tau) 
\psi_{\downarrow}({\mathbf r'},\tau') \right] \rangle  \, ,
\label{F-Matsubara}
\end{eqnarray}
where now $\psi_{\sigma}({\mathbf r},\tau) = \exp (K\tau)
\psi_{\sigma}({\mathbf r}) \exp (-K\tau)$,
$\psi_{\sigma}^{\dagger}({\mathbf r},\tau) = \exp (K\tau)
\psi_{\sigma}^{\dagger}({\mathbf r}) \exp (-K\tau)$,
and $T_{\tau}$ is the time-ordering operator for imaginary time $\tau$.

The Lehmann analysis for the normal function $G^{R}$ in the
broken-symmetry phase proceeds along similar lines
as for the normal phase~\cite{FW}.
The result is that (for a homogeneous system) the wave-vector and (real)
frequency Fourier transform can be
obtained by the spectral representation
\begin{equation}
G^{R}({\mathbf k},\omega) \, = \, \int_{-\infty}^{+\infty} \, d \omega' \,
            \frac{A({\mathbf k},\omega')}{\omega \, - \, \omega' \, + \, i
\, \eta}      \label{spectral-repres-G-R}
\end{equation}
$\eta$ being a positive infinitesimal. Here, the real and positive
definite \emph{spectral function\/} 
$A({\mathbf k},\omega) = - (1/\pi) \mathrm{Im}\, G^{R}({\mathbf k},\omega)$
satisfies the sum rule
\begin{equation}
\int_{-\infty}^{+\infty} \, d \omega \, A({\mathbf k},\omega) \, = \, 1
\label{sum-rule-G-R}
\end{equation}
for any given ${\mathbf k}$, as a consequence of the canonical 
anticommutation relation of the field operators.

A similar analysis for the Matsubara normal Green's function leads to
the spectral representation
\begin{equation}
G({\mathbf k},\omega_{s}) \, = \, G_{11}({\mathbf k},\omega_{s}) \, = \,
\int_{-\infty}^{+\infty} \, d \omega' \,
\frac{A({\mathbf k},\omega')}{i\omega_{s} \, - \, \omega'}  \,\, ,
\label{spectral-repres-G-Matsubara}
\end{equation}
in terms of the \emph{same\/} spectral function $A({\mathbf k},\omega)$ of
Eq.~(\ref{spectral-repres-G-R}), where
$\omega_{s}=(2s+1)\pi/\beta$ ($s$ integer) is a fermionic Matsubara
frequency and the diagonal Nambu Green's
function has been introduced.
The spectral representations (\ref{spectral-repres-G-R}) and
(\ref{spectral-repres-G-Matsubara}), together with
knowledge of the asymptotic behavior $G^{R}({\mathbf k},\omega) \sim
\omega^{-1}$ for large $|\omega|$, are sufficient to
guarantee that the retarded normal function is the correct analytic 
continuation of its Matsubara counterpart in the upper-half of the complex 
frequency plane \cite{FW}, in accordance with the Baym-Mermin 
theorem \cite{Baym-Mermin-61}.

The above Lehmann analysis can be extended to the anomalous function
(\ref{F-retarded}) as well.
One obtains
\begin{equation}
F^{R}({\mathbf k},\omega) \, = \, \int_{-\infty}^{+\infty} \, d \omega' \,
            \frac{B({\mathbf k},\omega')}{\omega \, - \, \omega' \, + \, i
\, \eta}           \label{spectral-repres-F-R}
\end{equation}
in the place of Eq.~(\ref{spectral-repres-G-R}). The new spectral
function $B({\mathbf k},\omega)$
vanishes for large $|\omega|$ but, in general, is no longer real and
positive definite.
[One obtains for $B({\mathbf k},\omega)$ the same formal expression~\cite{FW} 
for $A({\mathbf k},\omega)$ in terms of the eigenstates
$|n>$ of the operators $H$ and $N$, apart from the replacement of
$|\langle n'|\psi_{\uparrow}({\mathbf r}=0)|n \rangle|^{2}$
with
$\langle n|\psi_{\downarrow}({\mathbf r}=0)|n' \rangle$ $\langle n'
|\psi_{\uparrow}({\mathbf r}=0)|n\rangle$.]\cite{footnote-2}
It can then be readily verified that $B({\mathbf k},\omega)$ satisfies the sum 
rule
\begin{equation}
\int_{-\infty}^{+\infty} \, d \omega \, B({\mathbf k},\omega) \, = \, 0
\,\, , \label{sum-rule-F-R}
\end{equation}
which is again a consequence of the canonical anticommutation relation of the
field operators.
The above properties guarantee that $F^{R}({\mathbf k},\omega)$ vanishes
faster than $\omega^{-1}$ for large
$|\omega|$.

By a similar token, considering the Matsubara anomalous Green's
function leads to the spectral representation
\begin{equation}
F({\mathbf k},\omega_{s}) \, = \, G_{12}({\mathbf k},\omega_{s}) \, = \,
\int_{-\infty}^{+\infty} \, d \omega' \,
\frac{B({\mathbf k},\omega')}{i\omega_{s} \, - \, \omega'}  \,\, ,
\label{spectral-repres-F-Matsubara}
\end{equation}
where the off-diagonal Nambu Green's function has been introduced.
These considerations suffice again to guarantee that the retarded
anomalous function is the correct analytic
continuation of its Matsubara counterpart in the upper-half complex
frequency plane, in accordance with the
Baym-Mermin theorem \cite{Baym-Mermin-61}.

Finally, an additional sum rule for $B({\mathbf p},\omega)$ can be obtained by
using the relation
\begin{eqnarray}
&&\int_{-\infty}^{+\infty} d \omega \, B({\mathbf k},\omega) \, \omega  =
 i \, \int_{-\infty}^{+\infty} 
\frac{d \omega}{2 \pi} \, F^{R}({\mathbf k},\omega) \, \omega \, e^{- i
\omega \eta}              \nonumber \\
& &=  i \, \int \! d{\mathbf r} \, e^{- i {\mathbf k}\cdot{\mathbf r}} \,
\langle \left\{\frac{\partial \psi_{\uparrow}({\mathbf r},t=0^+)}{\partial
t},\psi_{\downarrow}(0)\right\}\rangle
\label{3rd-sum-rule-initial}
\end{eqnarray}
and exploiting the equation of motion for the field operator.
For the contact potential we are considering throughout this paper, we write
\begin{eqnarray}
\langle \left\{\frac{\partial \psi_{\uparrow}({\mathbf r},t=0^+)}
{\partial t},\psi_{\downarrow}(0) \right\}\rangle
&=& - v_{0} \delta({\mathbf r}) \langle
\psi_{\uparrow}({\mathbf r})\psi_{\downarrow}({\mathbf r})\rangle\nonumber\\
&=& - \delta({\mathbf r}) \Delta
\label{equation-of-motion}
\end{eqnarray}
in terms of the order parameter $\Delta$. The expression
(\ref{3rd-sum-rule-initial}) thus becomes:

\begin{equation}
\int_{-\infty}^{+\infty} \, d \omega \, B({\mathbf k},\omega) \, \omega \, =
\, - \, \Delta  \,\, . \label{3rd-sum-rule-final}
\end{equation}

\noindent
This constitutes a third sum rule for the spectral functions in
the broken-symmetry phase.


\begin{thebibliography}{99}
\bibitem{footnotelevin} Besides the difference mentioned in the text about the
relative order of performing the analytic continuation to real frequency
and of considering approximate expressions for the self-energy in the 
Matsubara representation, it is also worth mentioning some additional 
differences between our theory and the theory of Ref.~\onlinecite{levin98} 
based on a ``preformed-pair scenario'' below $T_c$. In our theory, the 
fluctuation 
propagator~(\ref{Gamma-solution}) is built on the BCS Green's functions 
(\ref{BCS-Green-function}) and is obtained via the inversion of a 2 $\times$ 2
matrix. In the theory of Ref.~\onlinecite{levin98}, the corresponding
propagator is instead built on a non-interacting Green's function 
${\cal G}_0$  and on a dressed Green's function with the 
functional form of Eq.~(\ref{G-11-strong-coupling}). We have verified that, 
in the extreme
strong-coupling limit (where the residual interaction between the composite 
bosons becomes irrelevant) and for temperatures $T\ll T_c$, the two theories 
give essentially the same results in the Matsubara representation. Differences
show up, however, at weaker couplings, when the interaction between composite
bosons matters. Specifically, our theory for the composite bosons reproduces 
the Bogoliubov results for point-like bosons with
 $\mu_B \neq 0$ and $n_0(T=0)\neq n$,
owing to the boson-boson interaction and the depletion of the condensate.
In the theory of Ref.~\onlinecite{levin98}, on the other hand, $\mu_B$ for
the composite bosons is taken to vanish for $T\le T_c$ and also $n_0(T=0)=n$ 
for all couplings. 
More generally for any coupling, in our theory (quantum) fluctuation 
corrections to mean-field quantities survive even at $T=0$ while in the 
theory of Ref.~\onlinecite{levin98} fluctuation corrections vanish 
identically at $T=0$.
\bibitem{footnote-2} Quite generally, the function $B({\mathbf k},\omega)$
introduced through the Lehmann representation (\ref{spectral-repres-F-R}) can
be cast in the form $B({\mathbf k},\omega) = [F^{A}({\mathbf k},\omega)
- F^{R}({\mathbf k},\omega)]/(2 \pi i)$, in terms of the \emph{advanced\/} 
($A$) and \emph{retarded\/} ($R$) ``anomalous'' functions.
In particular, when $F^{A}({\mathbf k},\omega) = 
F^{R}({\mathbf k},\omega)^{*}$ for \emph{real\/}
$\omega$, $B({\mathbf k},\omega)$ can be identified with 
$-(1/\pi) \mathrm{Im} F^{R}({\mathbf k},\omega)$.
This is the case for the ``anomalous'' retarded and advanced Green's functions
resulting from the approximate expression (\ref{G-12-Matsubara}) of the text 
with a \emph{real\/} order parameter $\Delta$.
\end{thebibliography}
\end{document}